%% file: main.tex
\documentclass[sigconf,nonacm]{acmart}
\AtBeginDocument{%
  }

\setcopyright{none}

\input{conf-format/imports}

\graphicspath{{./figures}{}}

\begin{document}

\title{ParTEETor: A System for Partial Deployments of TEEs within Tor}

\author{Rachel King}
\affiliation{%
  \institution{University of Wisconsin-Madison}
  \city{Madison, WI}
  \country{USA}}
\email{rachelking@cs.wisc.edu}

\author{Quinn Burke}
\affiliation{%
  \institution{University of Wisconsin-Madison}
  \city{Madison, WI}
  \country{USA}}
\email{qkb@cs.wisc.edu}

\author{Yohan Beugin}
\affiliation{%
  \institution{University of Wisconsin-Madison}
  \city{Madison, WI}
  \country{USA}}
\email{ybeugin@cs.wisc.edu}

\author{Blaine Hoak}
\affiliation{%
  \institution{University of Wisconsin-Madison}
  \city{Madison, WI}
  \country{USA}}
\email{bhoak@cs.wisc.edu}

\author{Kunyang Li}
\affiliation{%
  \institution{University of Wisconsin-Madison}
  \city{Madison, WI}
  \country{USA}}
\email{kli253@wisc.edu}

\author{Eric Pauley}
\affiliation{%
  \institution{University of Wisconsin-Madison}
  \city{Madison, WI}
  \country{USA}}
\email{epauley@cs.wisc.edu}

\author{Ryan Sheatsley}
\affiliation{%
  \institution{University of Wisconsin-Madison}
  \city{Madison, WI}
  \country{USA}}
\email{sheatsley@wisc.edu}

\author{Patrick McDaniel}
\affiliation{%
  \institution{University of Wisconsin-Madison}
  \city{Madison, WI}
  \country{USA}}
\email{mcdaniel@cs.wisc.edu}

\renewcommand{\shortauthors}{King et al.}

\begin{abstract}
  \input{conf-format/00.abstract}

\end{abstract}

\maketitle

\input{conf-format/01.introduction}
\input{conf-format/02.background}
\input{conf-format/03.system-overview}

\input{conf-format/04.methodology1}

\input{conf-format/05.methodology2}

\input{conf-format/06.evaluation}

\input{conf-format/07.discussion}

\input{conf-format/09.conclusion}

\bibliographystyle{ACM-Reference-Format}
\bibliography{refs}

\end{document}

%% file: conf-format/imports.tex
\usepackage{graphicx}
\usepackage{booktabs}
\usepackage{subcaption}
\usepackage{xcolor, pifont}
\usepackage{algorithm2e}
\usepackage{float}
\usepackage{comment}
\usepackage{bm}

\makeatletter\newenvironment{betteralgorithm}[1][!ht]{%
    \def\@algocf@post@ruled{%
        \kern\interspacealgoruled\hrule%
        height\algoheightrule\kern3pt\relax}%
    \def\@algocf@capt@ruled{under}%
    \setlength\algotitleheightrule{0pt}%
    \begin{algorithm}[#1]}{%
    \SetAlgoLined{}
    \end{algorithm}}
\makeatother

\SetKwInput{KwInput}{Input}
\SetKwInput{KwOutput}{Output} 
\SetKwProg{Fn}{Function}{:}{end}
\SetKw{Ret}{return}
\SetKwFunction{RelaySelection}{RelaySelection}

\newcommand{\shortsection}[2][.]{
    \vspace{2mm}\noindent\textbf{#2#1}}

\newcommand{\ParTEETor}{\texttt{ParTEETor}\xspace}

%% file: conf-format/00.abstract.tex
The Tor anonymity network allows users such as political activists and those under repressive governments to protect their privacy when communicating over the internet. At the same time, Tor has been demonstrated to be vulnerable to several classes of  deanonymizing attacks that expose user behavior and identities.   Prior work has shown that these threats can be mitigated by leveraging trusted execution environments (TEEs).  However, previous proposals assume that all relays in the network will be TEE-based---which as a practical matter is unrealistic.  In this work, we introduce \ParTEETor, a Tor-variant system, which leverages partial deployments of TEEs to thwart known attacks.  We study two modes of operation: non-policy and policy. Non-policy mode uses the existing Tor relay selection algorithm to provide users incident security.  Policy mode extends the relay selection algorithm to address the classes of attacks by enforcing a specific TEE circuit configuration.  We evaluate \ParTEETor for security, performance, and privacy.  Our evaluation demonstrates that at even a small TEE penetration (e.g., 10\% of relays are TEE-based), users can reach performance of Tor today while enforcing a security policy to guarantee protection from at least two classes of attacks.  Overall, we find that partial deployments of TEEs can substantially improve the security of Tor, without a significant impact on performance or privacy.

%% file: conf-format/01.introduction.tex
\section{Introduction}

Anonymity networks have existed for decades to provide users enhanced privacy when browsing the internet~\cite{zantout2011i2p, reiter1998crowds, freedman2002tarzan, danezis2003mixminion}. The onion routing network Tor~\cite{tor2004original} is a prominent example that allows users to access/host services anonymously and bypass censorship.
Entirely non-profit, Tor relies on people around the world volunteering their computing resources to host \textit{relays}, which route users' traffic through the network via unique, random paths known as \textit{circuits}. 
Currently, Tor has over 2 million users and over 6000 public relays in use~\cite{tormetrics}. 

While Tor enjoys widespread use, vulnerabilities in its infrastructure continue to emerge and threaten user privacy.
Attempts at exploiting protocol-level information (e.g., circuit identifiers) have been successful in deanonymizing users; known classes of attacks on Tor include collusion~\cite{replay_tagging_attack, hidden_service_attack}, fingerprinting~\cite{fingerprinting_attack}, circuit identifier exploitation~\cite{badapple_attack}, and bandwidth inflation~\cite{bandwidth_attack}. Ultimately, all of these attacks exploit the trust that is inherently placed in relays to safely route users' traffic. 
Developing a solution for effectively preventing such attacks is thus critical for the long-term success of Tor and the guarantees provided to users that rely on Tor.



Prior efforts like SGX-Tor~\cite{sgx-tor-archival} have used trusted execution environments (TEEs) to provide a means for trusting relays. However, the system assumes that every relay in the network is TEE-based. As the Tor network is entirely volunteer, and adopting TEEs necessarily requires software re-design and, for some relay operators, hardware changes, this greenfield assumption is at odds with practical considerations. Coordinating a transition to TEE-based relays among unassociated relay operators is infeasible, and thus this assumption is prohibitive. In practice, TEE-based solutions should tolerate partial rollout as relay operators begin adopting TEE-capable hardware and transitioning relay software. 
Thus, understanding the security guarantees of leveraging TEEs in partial (or incremental) settings will be pivotal to improving anonymity networks such as Tor.



In this paper, we introduce the \ParTEETor system which provides security in partial deployments of TEE-based relays in the Tor network.
Two modes of operation are provided: non-policy and policy-based. 
Non-policy mode allows users to benefit from TEE-based relays where they are available using the current Tor relay selection algorithm. 
Policy mode allows users to mandate specific TEE circuit configurations to guarantee protection from different classes of attacks.


As an initial means of evaluating partial deployments, we create a taxonomy of classes of attacks on Tor. We then map those classes to TEE circuit configurations that mitigate them. The identified configurations are designated as \textit{security policies}. We introduce an \textit{extended relay selection algorithm} that selects circuits that meet a client-specified security policy. 


Next, we develop a simulation of \ParTEETor using Python to evaluate its security, performance, and privacy over four realistic \textit{deployment scenarios} of TEEs. 
We evaluate security as the general TEE protection circuits receive with no security policy enforced. 
To evaluate performance, we use measurements provided by the Tor directory consensus for expected bandwidth of circuits when enforcing a security policy. 
Privacy is evaluated as the resulting reduction in circuit availability when enforcing a security policy.

We find that non-policy mode offers protection from at least two classes of attacks to more than $50\%$ of circuits if at least $20\%$ of relays are TEE-based.  
A possible consequence of policy mode is increased congestion at TEE-based relays if there is a limited penetration of TEEs.
However, we still find that with only 10\% of TEE-based relays, users reach an expected bandwidth of 8016.1 KB/s, meeting performance seen in Tor today while obtaining protection from two classes of attacks. 
While enforcing a security policy also reduces the space of available circuits, the reduction in this space still meets the privacy guarantees observed in former versions of Tor. 

Our work shows that the use of TEEs in Tor is both practical and effective. We hope that our work encourages relay operators to begin taking concrete steps towards using TEEs to reinforce the long-term success of Tor and anonymity networks.

%% file: conf-format/02.background.tex
\section{Background}\label{background}
\shortsection{Tor Network}
Tor~\cite{tor2004original} is a volunteer-based anonymity network that uses onion routing~\cite{onion_routing} to conceal a user's internet activities. 
The main components of Tor are \textit{clients}, \textit{relays}, \textit{directory authorities}, and \textit{onion services}. Implemented as an overlay network, Tor encapsulates client TCP streams in multiple layers of encryption and routes them through several intermediate servers between a source (client) and destination (website). No component is aware of both the source and destination, thus allowing clients to remain anonymous to the wider internet (i.e., to relays, web servers, and other third-parties).

A user's Tor client establishes a \textit{circuit} to send data by choosing three (or more) relays during \textit{relay selection}. Ephemeral, pairwise encryption keys are negotiated between the client and each relay. To anonymize data for transmission, the client divides the data into fixed-size \textit{cells} and successively encrypts every cell under each key, beginning with the last relay's key, followed by the middle, then the first. During transmission, each relay decrypts the cell to remove their layer of encapsulation before forwarding it to the next relay in the circuit. A relay is therefore only aware of the identities of components immediately prior and after itself in the circuit.

Directory authorities are a subset of ten relays in the network that are trusted to maintain the overall state of the network. Their central responsibility is upholding the \textit{directory consensus document}, a document containing key information about each relay (e.g., public keys, bandwidth, flags, etc.) that clients query during relay selection. The directory authorities scan the network periodically to update the consensus document. 

Onion services (formerly known as hidden services) are special sites and services only accessible through Tor. The key components of onion services are the \textit{introduction points}, which are anonymous contact points of the site hosts, and \textit{rendezvous points}, which connect anonymous circuits from the client and the site host. 
Clients first communicate with the introduction point of the onion service to identify a rendezvous point.
Onion services then have their inbound traffic routed through a circuit to the rendezvous point. Clients also establish a circuit of their own to the rendezvous point.

\shortsection{Trusted Execution Environments}
Trusted execution environments (TEEs) are hardware-based security primitives that protect the confidentiality and integrity of code and data running in untrusted environments (e.g., on outsourced servers)~\cite{mckeen2013innovative,ngabonziza_trustzone_2016,kaplan_amd_2016}. TEEs protect confidentiality primarily through \textit{access-mediation} to protected memory regions containing sensitive code and data, and additional \textit{CPU modes} that restrict the type and scope of operations that can be performed by code running either within the context of the TEE or outside of it. In effect, this ensures complete isolation from untrusted third-parties while processing sensitive data. \textit{Attestation} capabilities are also provided to allow clients to verify the integrity of the code running inside the TEE. This grants clients assurance of the behavior of the code running on machines they do not physically manage.
TEEs have begun to see wide adoption for securing cloud-based web services~\cite{graphene}, but also for improving the security of Tor.

\shortsection{SGX-Tor}
The SGX-Tor system demonstrated that TEEs provide an effective means to defend against several longstanding threats to Tor~\cite{sgx-tor-archival}. This was realized by porting the Tor software to run within a TEE.
This ensures that sensitive data is kept secret and functions processing sensitive data cannot be tampered with. To enable communication with the outside world, the host that the TEE runs on acts as a proxy and forwards messages to endpoints specified by the TEE (e.g., to clients and other relays in Tor). Any data that requires persistence on the relay can then be sealed using a private key burned into the processor hardware and only accessible by the TEE code~\cite{sgx}. More broadly, the confidentiality, integrity, and attestation capabilities of the TEE thus ensure the trustworthiness of the relays in the network, preventing an otherwise untrustworthy relay from leaking sensitive information like circuit IDs, cell commands, router descriptors, or private encryption keys used for communication with clients. This effectively reduces the threat surface within Tor to only network-level adversaries.

%% file: conf-format/03.system-overview.tex
\section{System Overview}
SGX-Tor~\cite{sgx-tor-archival} laid the foundation for integrating TEEs into Tor.
However, it was limited by the assumption that every Tor client and relay has a TEE. 

This greenfield assumption is impractical as it implies the entire network will be replaced, instead of allowing for compatibility of both TEE and non-TEE relays. While many systems already have the hardware to support TEEs, adoption still requires software re-design. For some relay operators, adoption could also mean hardware changes. Coordinating a transition to all TEE-based relays among unassociated relay operators is infeasible, and thus this assumption is prohibitive.
 
To stimulate adoption of TEEs by the broader Tor community, our central goal is therefore to construct a system that leverages partial deployments of TEE-based relays to enable users security, performance, and privacy.

\begin{figure}[t!]
    \centering
    \resizebox{\columnwidth}{!}{
        \includegraphics{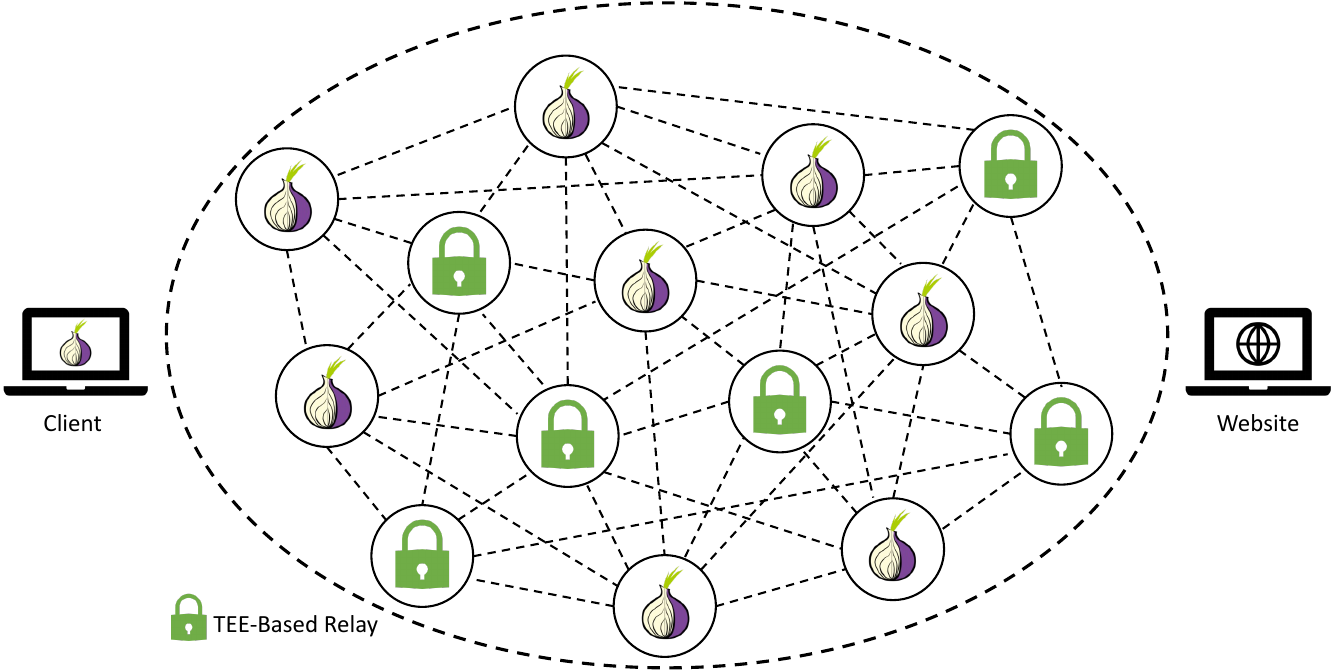}
    }
    \caption{\ParTEETor, a partial deployment of TEE-based relays in the Tor network.}
    \label{fig:parteetor}
\end{figure}

\shortsection{ParTEETor}
Based on the existing Tor design, the \ParTEETor system (\autoref{fig:parteetor}) integrates TEEs in some relays---varying from very few to nearly all (See Section~\ref{evaluation}). 

While one might expect to need hardware changes in order to use \ParTEETor, this is not necessarily the case. 
Implementing TEEs is standard practice now for general purpose CPUS. Popular manufacturers like Intel, ARM, AMD, and Apple all utilize TEEs in their CPUs~\cite{sgx_processors, sev_processors, trustzone2009, trustzone2017, apple_enclave}. Therefore, many relay operators will only need to install a binary which utilizes their already existing TEE in order to use \ParTEETor.

Like SGX-Tor, the relay code of TEE-based relays is moved into the TEE, thereby providing data and operation integrity and confidentiality guarantees.  Moreover, attestation is used to ensure the integrity of the implementation (code).
Thus, when a relay wishes to be available to clients as TEE-based, it must first be attested by the directory authorities, and re-attested at some periodicity to ensure the ongoing integrity of the relay. 

\ParTEETor operates in one of two modes: non-policy and policy (See Section~\ref{methodology2}). 
The non-policy mode is a straightforward application of Tor with the addition of TEEs being adopted by select existing relays. 
The relay selection algorithm in this mode does not change, allowing for circuits to incidentally benefit from the availability of TEE-based relays in the network. This mode represents the case where no changes to the client are required, as all changes to behavior exist solely in the TEE-based relays.


\ParTEETor's policy mode extends non-policy mode by incorporating security policies into the relay selection algorithm. When a client connects to the network, it will select its circuit based on its desired security policy, which defines which relays in a circuit are TEE-enabled (See Section~\ref{methodology1}). The security policy enables users to choose their protection status from each class of attack based on the perceived threats. This mode requires updating the client to impose the selection of TEE-based relays in circuits.

\shortsection{Threat Model and Assumptions} We assume a similar threat model to other TEE-based systems~\cite{graphene,sgx-tor-archival}. Adversaries are exploiting the required trust placed on relays to safely route traffic through the network. 
An adversary may be a compromised entry, middle, or exit relay in the network. They may attempt to corrupt or extract private information from the relay, such as circuit IDs or cell commands. However, we assume that the confidentiality and integrity of any code and data processed within the TEE is protected. We do not trust software running outside of the TEE (i.e., the operating system acting as a proxy for the TEE). We also assume that users trust their Tor client and therefore do not consider malicious client attacks~\cite{demultiplex_attack, congestion_attack, jansen_sniper_2014}. We assume that all directory authorities are trustworthy, as these relays are operated by trusted parties in the Tor community. 
We note the limitations of TEEs in Section \ref{discussion}.

%% file: conf-format/04.methodology1.tex
\section{Mapping Attacks to Circuit Protection}\label{methodology1}

In order to understand what security policies \ParTEETor needs to support, we need to understand the relationship between TEE-based relays and their placement in circuits to 
protect against the known classes of attacks on Tor users (\autoref{tab:circuit-mappings}).
To do this, we present a security analysis on the known classes of attacks in Tor. 
These attacks were identified by prior work as the universal examples for each known attack class~\cite{sgx-tor-archival}. 


As not all attacks require collusion of every relay in a circuit, ensuring a TEE-based relay is present in every position of a circuit would not be necessary for protection from each attack.
Prior work provided a preliminary analysis on the minimum requirement of TEE-based relays~\cite{sgx-tor-archival}; we expand on their efforts by detailing how the specific TEE placement in a circuit provides protection against each attack. We breakdown each attack to identify the specific position (i.e., distinct relays in a circuit) where TEEs are required in a circuit to ensure mitigation.

\subsection{Replay Attack}
\begin{figure}[!h]
\centering
\centerline{\includegraphics[scale=0.5]{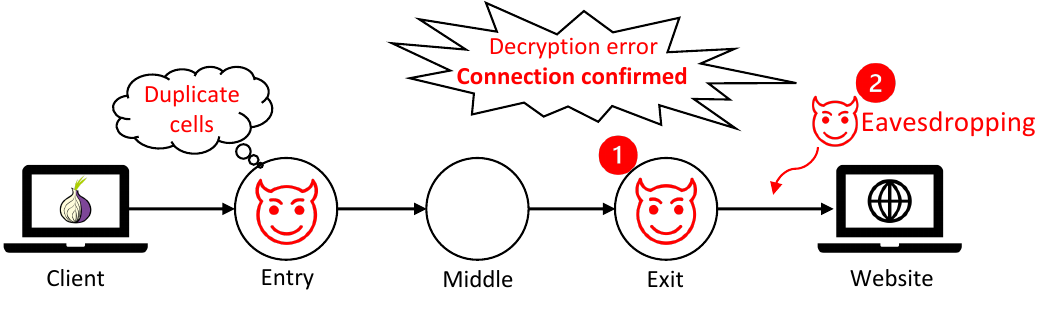}}
\caption{Replay Attack Scenario}
\label{fig:replay_attack}
\end{figure}

Pries \textit{et al.} propose the Replay Attack~\cite{replay_tagging_attack}, where an adversary duplicates Tor cells, causing decryption errors. As Tor uses AES-CTR for cell encryption, each relay and client maintain a counter when encrypting and decrypting along a circuit. Duplicating a cell will result in the counter being additionally incremented at the relays decrypting it, which causes the client and the relays' counters to be out of sync. This will then cause a decryption error at either edge of the circuit. 

This attack can take two approaches, as shown in \autoref{fig:replay_attack}: both the entry and exit relays to be malicious, or the entry is malicious and an adversary is eavesdropping on the connection between the exit relay and the destination. 
The initial premise is a malicious entry relay duplicates a relay cell before forwarding it down the circuit. Once the duplicated cell reaches the exit relay, the integrity check will fail and cause the circuit to be torn down. 
A malicious exit relay working with the entry can confirm the user and the destination, as this error was forced by the entry for the exit to recognize. A malicious eavesdropper, on the other hand, will notice the TCP stream being cut off unexpectedly, confirming the connection between the entry and the eavesdropper.

\input{figures/attacktoconfigmapping}

\shortsection{Protection}
The Replay Attack~\cite{replay_tagging_attack} relies on malicious relay behavior. Specifically, the entry relay must duplicate the target cell to cause the decryption error. 
The guarantee of source code integrity with a TEE prevents this behavior. 
Protection from this attack requires a TEE-based entry relay. It is not necessary to require both a TEE-based entry relay and exit relay. If the entry relay is TEE-based, we can guarantee this relay will not duplicate any cells, meaning any adversary at the exit relay will have no decryption error.

Considering the variation of the attack where the entry relay is malicious with an eavesdropper on the exit-to-destination link, a TEE cannot prevent this eavesdropper. This is the reasoning behind requiring the \textbf{entry} relay to have a TEE and not the exit: preventing the duplication of the cell from ever taking place at the entry, thereby impeding the eavesdropper from noticing any unexpected dropped connection.

\subsection{Onion Services Attack}

\begin{figure}[!h]
\centering
\centerline{\includegraphics[scale=0.5]{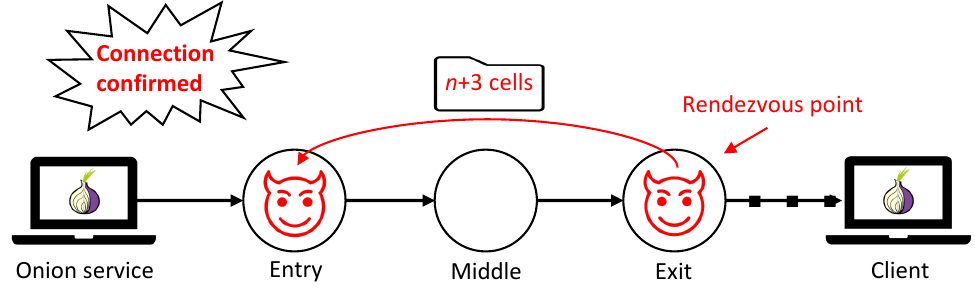}}
\caption{Onion Services Attack Scenario}
\label{fig:hidden_services_attack}
\end{figure}

Biryukov \textit{et al.} propose an onion services attack~\cite{hidden_service_attack} which relies on a malicious entry or middle relay, and rendezvous point to reveal the location of onion services, as shown in \autoref{fig:hidden_services_attack}.
The adversary intends to confirm they are running the entry relay for an onion service by recognizing a pattern of cells sent by the rendezvous point. When the adversary requests to be introduced to the onion service as a standard Tor user, they provide their malicious rendezvous point to the onion service. The onion service then constructs a circuit to the rendezvous point, where the adversary sends a known quantity $n$ (e.g. 50) of \textit{padding} cells down the circuit, followed by a \textit{destroy} cell.

If the adversary is the entry relay in this circuit to the rendezvous point, they will receive $n+3$ cells in total, two \textit{extended} cells from circuit establishment, $n$ \textit{padding} cells, then one \textit{destroy} cell. This scenario confirms the adversary is the entry relay of the onion service, allowing them to reveal the identity of the onion service as the hop prior to them.
If the adversary is only the middle relay, it will receive $n+2$ cells total (one less \textit{extended} cell), confirming the hop prior is the entry of the circuit (we will disregard this scenario from here on, as compromising the entry relay following this scenario would require additional attacks outside of this).


\shortsection{Protection}
The Onion Services Attack~\cite{hidden_service_attack} requires the rendezvous point to maliciously send padding cells down the circuit. However, the integrity guarantee of TEEs prevents this behavior.
The attack also requires a malicious entry in order to recognize the padding cells and count the total cells sent. Thus, a TEE-based entry relay alone does not prevent this attack, as a network-level adversary can still observe the number of cells sent via analysis of the quantity and size of IP packets. 
To prevent an adversary from inferring this link, the act of transmitting extra padding cells must be prevented. This requires a TEE-based relay acting as the rendezvous point, which translates to the \textbf{exit} relay in the circuit.




\subsection{Fingerprinting Attack}

\begin{figure}[!h]
\centering
\centerline{\includegraphics[scale=0.5]{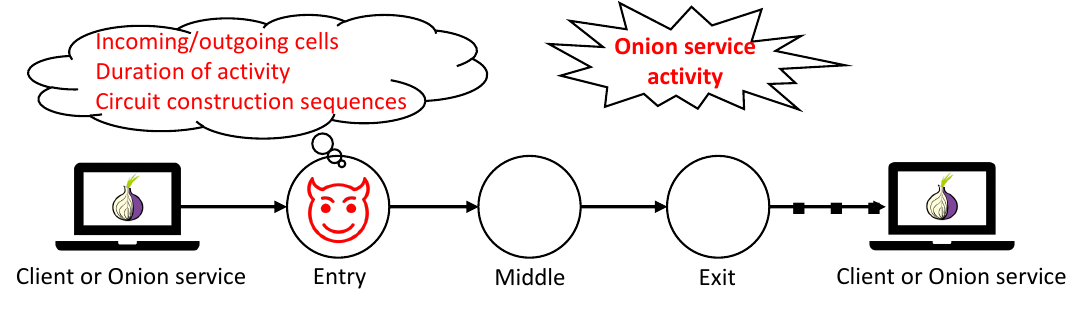}}
\caption{Fingerprinting Attack Scenario}
\label{fig:fp_attack}
\end{figure}

Kwon \textit{et al.} propose a fingerprinting attack~\cite{fingerprinting_attack} which is able to exploit circuit level identifying information to reveal if a given user is visiting an onion service, or the location of the onion service itself.
This attack, shown in \autoref{fig:fp_attack}, requires the adversary to be acting as the entry relay.

The adversary makes note of three different properties of the traffic to and from onion services: incoming and outgoing cells, duration of activity, circuit construction sequences. Based on these properties, introduction point circuits are first sought out, meaning circuits between a client and introduction point for an onion service. Once evidence of these circuits is found, the adversary monitors the users of these circuits further to determine rendezvous point circuits, either between a client or an onion service. The activity monitored can effectively determine if the adversary is an entry relay for an onion service or a user visiting an onion service. In the event the relay is acting as the entry for an onion service, the location of the onion service is now identified.


\shortsection{Protection}
The Fingerprinting Attack~\cite{fingerprinting_attack} requires the entry relay to be malicious to recognize patterns of cells being sent across circuits. Additionally, the authors claim the attack can be implemented by someone eavesdropping on the connection between the user and the entry relay. This attack is passive in that it only requires observing circuit identifying information, such as circuit IDs and number and sequences of cells, to distinguish different circuits. TEEs' guarantee of confidentiality prevents this information from being revealed to a malicious entity on the relay. 
A malicious host or entity eavesdropping on the connection will only see IP packets being sent. 

Protection from this attack requires a TEE-based \textbf{entry} relay. This ensures the entry cannot recognize what traffic corresponds to specific circuits.
An important note is that TEEs only reduce the adversary to a network level. This attack could still be possible via analysis of IP packets in an attempt to distinguish circuits.



\subsection{Bad Apple Attack}

\begin{figure}[!h]
\centering
\centerline{\includegraphics[scale=0.5]{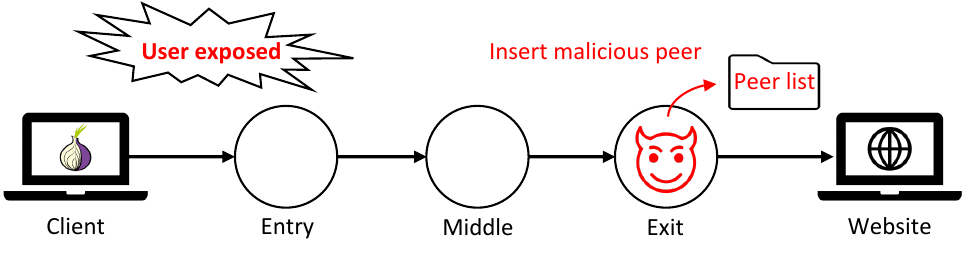}}
\caption{Bad Apple Attack Scenario}
\label{fig:bad_apple_attack}
\end{figure}

Tor multiplexes multiple TCP streams over one circuit. In the event a user is visiting a website, which may require multiple streams to fetch all the objects, these streams will be sent over the same circuit. Subsequently, if an exit relay is able to identify the source for one of the streams, it now knows the source of all the other streams along that circuit. The Bad Apple Attack~\cite{badapple_attack}, shown in \autoref{fig:bad_apple_attack},
exploits this design choice. 

Targeting users of peer-to-peer (P2P) file sharing applications like
BitTorrent, this attack requires the adversary to host a malicious exit relay, monitor users of P2P applications, and host a malicious peer for the applications. 
Blond \textit{et al}'s attack exploits the fact that 70\% of users accessing BitTorrent only use Tor to request peers, then connect directly to the peer via TCP. 

When a user's circuit contains the malicious exit relay, and the user is requesting a list of peers to contact, the exit relay can modify the returned list to include their malicious peer. Then, when the user connects to the malicious peer outside of Tor, the user exposes their IP address (by design of P2P applications).
The adversary operating the relay can then match the traffic to its clearnet peer to clients it sends data back to as an exit relay.
Furthermore, the source of all other multiplexed streams of this particular exit relay's circuit are now exposed.

\shortsection{Protection}
The Bad Apple Attack~\cite{badapple_attack} protection is straightforward. This attack relies on the exit relay in a circuit to act maliciously. Moreover, no collusion is required for this attack, but the knowledge of circuit level information such as circuit and stream IDs is required to be able to distinguish between different circuits and streams. 
TEEs protect against the malicious relay behavior through their integrity guarantee. TEEs also hide circuit identifying information such as circuit and stream IDs. This prevents the adversary from recognizing different circuits through IDs. For these reasons, requiring the \textbf{exit} relay in a circuit to be TEE-based ensures protection under TEE capabilities. 
An important note is that the adversary is reduced to a network level, as they can still attempt to recognize different circuits and streams through IP packet inspection.

\subsection{Bandwidth Inflation Attack}

\begin{figure}[!h]
\centering
\centerline{\includegraphics[scale=0.5]{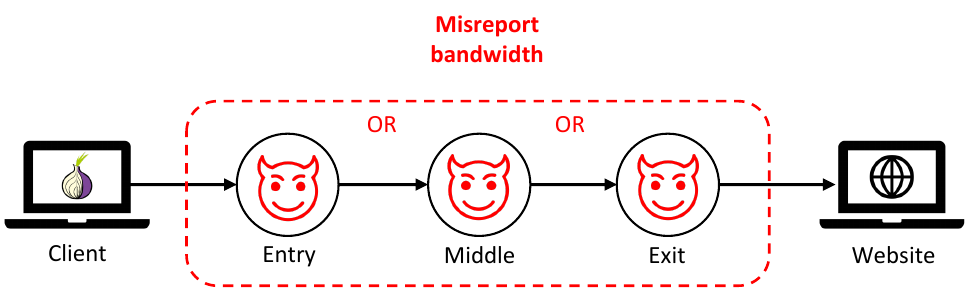}}
\caption{Bandwidth Inflation Scenario}
\label{fig:bandwidth_inflation}
\end{figure}

Relays are selected for circuits weighted proportionally to their bandwidth. This creates an incentive for having higher bandwidth, as that relay is now more likely to be used in circuits. From an adversarial perspective, this is beneficial in their task of controlling entry and exit relays of a circuit~\cite{bandwidth_attack}.
Tor adopts the approach of bandwidth scanners to validate the reported bandwidth. A portion of the directory authorities are bandwidth authorities and will periodically scan the bandwidth via bidirectional probing.
The published bandwidth of the relay is the median of at least three of the bandwidth authorities' measurements~\cite{bandwidth_scanner}. 

Bandwidth scanning is an improvement in preventing misreporting, shown in \autoref{fig:bandwidth_inflation}, but doesn't defeat the attack entirely. As relays are able to recognize the directory authorities that scan for bandwidth, relays can provide more bandwidth to these streams by throttling the bandwidth they allow for the rest of its streams. This can effectively convince the directory authorities that the relay is capable of higher bandwidth which in turn increases the probability it will be chosen for circuits, as shown by Biryukov \textit{et al.}~\cite{hidden_service_attack}.

\shortsection{Protection}
Relays using TEEs self-report their bandwidth~\cite{sgx-tor-archival} in place of scanning. With this, any manipulation to the bandwidths being reported to the directory authorities would mismatch the bandwidth reported by the trusted TEE relay.
As this attack is specific to each relay, protection requires \textbf{all (entry, middle, and exit)} relays in a circuit to be TEE-based. 

Requiring the entry and exit relays in a circuit to be TEE-based would be an effective mitigation if considering bandwidth inflation as a means to leverage more attacks. Most attacks discussed require collusion, except for Fingerprinting~\cite{fingerprinting_attack} (malicious entry relay) and Bad Apple~\cite{badapple_attack} (malicious exit relay). Requiring all relays in a circuit to be TEE-based would then not be necessary.

%% file: figures/attacktoconfigmapping.tex
\begin{table*}[!t]
    \centering
    \resizebox{\textwidth}{!}{
        \begin{tabular}{|l|l|l|l|}
        \hline
        \textbf{Attack} & \textbf{Adversary Relays} & \textbf{Adversarial Goal} & \textbf{TEE Requirement} \\ \hline
        \textbf{Replay} & Entry and Exit & Deanonymize users & Entry       \\
        \textbf{Onion Services} & Entry and Exit & Deanonymize onion services & Exit           \\
        \textbf{Fingerprinting} & Entry & Deanonymize users and onion services & Entry       \\
        \textbf{Bad Apple} & Exit & Deanonymize users & Exit          \\
        \textbf{Bandwidth Inflation} & Entry, Middle, and Exit & Increase relay's usage in circuits & Entry, Middle, and Exit \\ \hline
        \end{tabular}
    }
    \caption{Required TEE placement to mitigate attacks against Tor.}
    \label{tab:circuit-mappings}
\end{table*}

%% file: conf-format/05.methodology2.tex
\section{\ParTEETor}\label{methodology2}

\ParTEETor is driven primarily by two components: a client's \textit{security policy} and the \textit{extended relay selection algorithm}.  Additionally, how TEEs are deployed within the system (i.e., the  \textit{deployment scenario}) will have significant impact on its effectiveness. We discuss each aspect of the system and its use below.

\subsection{Circuit Security Policy}
\sloppy A client's security policy specifies which attacks a circuit must provide protection against. Concretely, it describes what TEE-based relays must be present (and how they must be arranged) within a circuit to provide protection. Our mapping from \autoref{tab:circuit-mappings} prescribes five policies: $\{\emptyset\}$, $\{Entry\}$, $\{Exit\}$, $\{Entry, Exit\}$, and $\{Entry, Middle, Exit\}$. For non-policy mode, a security policy is therefore empty.  We evaluate the performance, security, and privacy consequences of enforcing the policies in realistic deployments in the following sections.

\begin{betteralgorithm}[t!]
\Fn{\RelaySelection($G=(V,E), R=(P, T)$)}{
    $circuit$ = [ ]\;
    \For{position, TEEreq $ \in R$}{
        relaylist = \{ \}\;
        \For{$v \in V$}{
            \If{position $\in $ v.positions}{ 
                \If{v.TEE or not TEEreq \textcolor{blue}{\tcp{Security Policy Extension}}}{ 
                    add $v$ to $relaylist$\;
                }
            }
        }
        totalBW = $\sum_{\forall r \in relaylist} r.bandwidth$\;
        select relay $r$ with probability $\frac{r.bandwidth}{totalBW}$\;
        add $relay$ to $circuit$\;
    }
    \Ret{$circuit$} 
}
\caption{Bandwidth weighted relay selection for circuits. $G=(V,E)$ is the graph representing the Tor network and $R$ represents the configuration for the circuit, $position$, which contains the relay types (default is Entry, Middle, Exit), and security policy of TEE requirements for the circuit, $TEEreq$. $v.positions$ is the individual relay's circuit position capabilities, and $v.TEE$ is its TEE status.}
\label{alg:relayselection}

\end{betteralgorithm}

\subsection{Extended Relay Selection Algorithm}
Our extended relay selection algorithm adapts the relay selection algorithm currently used by Tor to enable clients to find policy-compliant circuits. In the base algorithm~\cite{torspec_git}, relays are weighted solely by their available bandwidth when being selected for use in circuits. In our extended algorithm, client select circuits in one of two ways. In non-policy mode, the algorithm simply falls back to the base algorithm. However, in policy mode, it additionally considers both the client's security policy and the availability of TEE-based relays in the network (published by the directory authorities in the directory consensus document). An overview of the algorithm is shown in~\autoref{alg:relayselection}. 
It works by iteratively searching the set of all relays to identify candidate TEE-based relays to use at each position in the circuit being constructed. Relays are similarly weighted by their available bandwidth during selection (line 14).

\begin{figure*}[!t]
    \centering
    \resizebox{\textwidth}{!}{
        \includegraphics{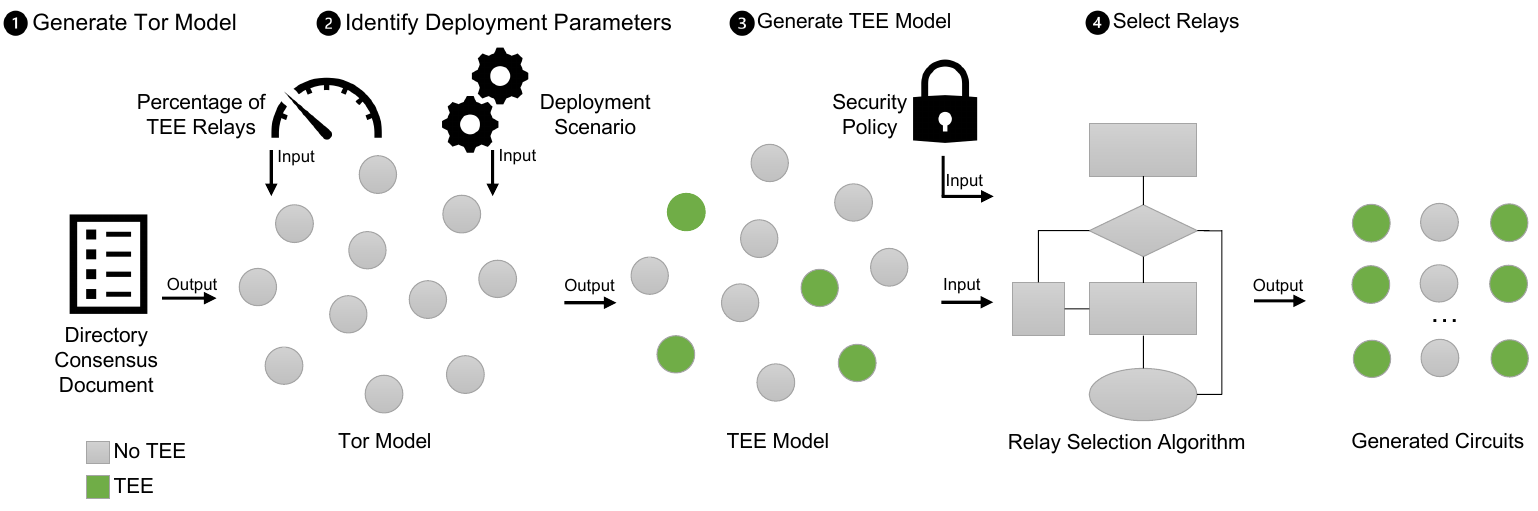}
    }
    \caption{Pipeline of events for modeling partial deployments of TEEs in \ParTEETor. A Tor network model is generated using relay data from the directory consensus. Relays are probabilistically assigned to be TEE-based for each deployment scenarios. Using Tor's relay selection algorithm with an additional parameter for TEE security policy, circuits are generated for analysis of their security and performance properties.}
    \label{fig:pipeline}
\end{figure*}

\subsection{Deployment Scenarios}\label{deployments}
The availability of TEE-based relays in the Tor network (and their arrangement within a given circuit) directly impacts user security, performance, and privacy.
To understand the efficacy of partial deployments, we introduce the notion of deployment scenarios. A deployment scenario describes a strategy for how relay operators could realistically begin upgrading their relays to run TEE-based hardware and software. We introduce four deployment scenarios that broadly capture such strategies: \texttt{Random}, \texttt{Bandwidth Weighted}, \texttt{Inverse Bandwidth Weighted}, and \texttt{Circuit-Position Weighted}. The first three allow a relay operator to decide which relays to upgrade by assigning a probability of being chosen for upgrade to each relay (either randomly or based on general relay characteristics like available bandwidth). The last allows a relay operator to more intelligently select relays for upgrade by assigning select probabilities to relays occupying distinct circuit positions.

\shortsection{Random Deployment}
\texttt{Random} deployment weights $w$ for each relay uniformly to represent a fully randomized selection. Explicitly, $w=1/n$ where $n$ is the number of relays in the network. No relay-specific characteristics are taken into account.

\shortsection{Bandwidth Weighted Deployment}
\texttt{Bandwidth Weighted} deployment assigns weight $w$ based on its bandwidth. This weight remains fixed. This approach is much like, and motivated by, the relay selection of Tor.
Relays with higher bandwidth are more likely to be chosen for circuits because they can withstand more traffic. Considering this, these relay operators have more incentive to adopt TEEs. 
We consider this approach a best case scenario, in that requiring TEE-based relays will likely increase a user's performance.

\shortsection{Inverse Bandwidth Weighted Deployment}
\texttt{Inverse Bandwidth Weighted} deployment assigns weight $w$ based on the inverse of a relay's bandwidth. This weight remains fixed. 
This approach is the opposite of the \texttt{Bandwidth Weighted} deployment. 
Our motivation for this deployment is to understand the worst case scenario in terms of performance when requiring TEE-based relays, as now all of the relays with lesser-bandwidth will have a higher likelihood of being assigned to be TEE-based.

\shortsection{Circuit-Position Weighted Deployment}
\texttt{Circuit-Position Weighted} deployment consists of four deployment distributions: Entry, Exit, Entry-Exit, and Entry-Middle-Exit.
Each distribution selects relays specifically with respect to their circuit position capabilities. Therefore, each deployment is dependent on three weights: $w_e$ = TEE-based entry relays, $w_m$ = TEE-based middle relays, and $w_x$ = TEE-based exit relays. Each of these weights represent the percentage of which are TEE-based, with respect to the total number of relays of that type.

In terms of the capabilities of relays with respect to their position in a circuit, every relay in the network is middle-capable. All entry relays can be in both the entry and middle positions of a circuit, and all exit relays can be in middle and exit positions. Additionally, some relays are both entry and exit-capable. Therefore, when considering these weights, under distributions Entry-Exit, and Entry-Middle-Exit, selection of relays may overlap. For example, selecting 10\% of entry relays to be TEE-based, then selecting 10\% of exit relays to be TEE-based could result in actually 15\% of entry relays being TEE-based because half of the exit relays chosen could also be entry-capable. 

As each distribution assigns TEE-based relays according to their circuit position capabilities, weight assignments are dependent on the specific distribution:
\begin{itemize}
    \item Entry: only vary the entry-capable relays, $w_e$
    \item Exit: only vary the exit-capable relays, $w_x$
    \item Entry-Exit: vary the entry-capable relays, $w_e$, and the exit-capable relays, $w_x$
    \item Entry-Middle-Exit: vary all relays, $w_e$, $w_m$, $w_x$
\end{itemize}

These deployment distributions provide insight into the impact each type of relay has on the network. They may be useful in assessing the trade-offs between relay cost and expected security/performance improvement (e.g., deciding to upgrade entry relays first, as they provide significant bandwidth to the network).

%% file: conf-format/06.evaluation.tex
\section{Evaluation}\label{evaluation}
We evaluate \ParTEETor via simulation of the real Tor network.  The experiments here seek to answer the following questions:
\textit{(1) How many circuits benefit from the addition of TEEs under the existing relay selection algorithm?} 
\textit{(2) How much congestion is present in circuits when enforcing TEE requirements under partial deployments?} 
\textit{(3) What is the reduction in availability of circuits when enforcing TEE requirements under partial deployments?}
We begin by briefly introducing the simulation framework and experimental evaluation.

\shortsection{Simulation Framework}
The \ParTEETor simulation models its behavior under different partial deployments of TEEs in the Tor network (\autoref{fig:pipeline}). We model the network as a graph~\cite{jansen2012methodically}, where each relay uses data from the advertised Tor network (i.e., IP address, expected bandwidth) as well as a TEE status representing whether the node has a TEE or not. 
Under each deployment scenario, we probabilistically assign relays to be TEE-based. For non-policy mode, we apply the existing Tor relay selection algorithm to simulate the behavior of selecting relays for circuits. 
We extend the algorithm by including an additional parameter for the security policy of the circuits for policy mode. We then generate circuits for two sets of experiments.

\input{figures/security}

\subsection{Experiment Setup}
Our \ParTEETor simulator is written in Python, using the \texttt{networkx} library to model relations between relay nodes. 
Simulations were performed on a Mac M1 CPU with 16 GB of ram and 3.2 GHz max clock speed.
We use a directory consensus document published on February 26, 2023~\cite{stempython} to generate our graph. In total, there are 6356 nodes, with 3179 having entry capability, and 1668 having exit capability. 849 nodes overlap in the categories, being both entry- and exit-capable.
We run our extended relay selection algorithm to generate 1000 potential circuits per trial for each security policy. We run each trial 10 times and take the mean to normalize our results. 

For \texttt{Random}, \texttt{Bandwidth Weighted}, and \texttt{Inverse Bandwidth Weighted} deployments, we denote $p$ as the percentage of total TEE-based relays in the network. We evaluate $p$ for the range from 1\% to 100\%. Depending on the specific deployment, the distribution of TEEs will vary. 
Recall that $w$ is the probability a relay will be TEE-based and is specific to each deployment (see \autoref{deployments}). 

As \texttt{Circuit-Position Weighted} deployment specifically targets relays with respect to their circuit position capabilities, $p$ is not an experimental parameter. Instead, weights $w_e$, $w_m$, and $w_x$ are iterated through as experimental parameters for each TEE distribution of this deployment. When a weight is a variable in the deployment, we evaluate it over a range from 1\% to 100\%.

\input{figures/security_2.tex}

\shortsection{Evaluation Metrics}
Our security metric is defined as the number of circuits generated that satisfy each security policy, when no policy is provided. 
This reflects the security implications of \ParTEETor in non-policy mode, where TEE-based relays are present in the network and the existing relay selection algorithm is not extended.

Our performance metric is defined as the expected median bandwidth of circuits generated with a security policy enforced. This reflects \ParTEETor in policy mode, where TEE-based circuit requirements are enforced through the extended relay selection algorithm.
We evaluate performance in terms of expected bandwidth, as this is a standard metric for performance in the Tor network~\cite{tuneup, panchenko_improving_2012}. Additionally, the cost of requiring TEE-based relays in circuits is increased congestion, which we can study via the expected bandwidth. 
To compute the expected bandwidth of a circuit, we consult the directory consensus document for expected bandwidth measurements of all relays, as measured by the bandwidth directory authorities. Next, we divide each relay's expected bandwidth by the number of circuits it is a part of. The expected bandwidth of a circuit is then the minimum bandwidth of all relays in the circuit. 

We do not consider latency in our performance evaluation, as it is not a factor in the relay selection algorithm. Due to the minimal delays (3.9\%) incurred by TEEs in end-to-end performance of Tor~\cite{sgx-tor-archival}, our simulation assumes zero overhead of TEEs. 

Our privacy metric is defined as the number of unique circuits that are possible under security policies when there is a limited TEE percentage. This reflects \ParTEETor in policy mode and presents the reduction in space of circuits, as specifying a policy reduces the size of the network to TEE-based relays.

\subsection{Security of \ParTEETor in Non-Policy Mode}



In order to understand the security implications of \ParTEETor in non-policy mode, we investigate what TEE coverage circuits can achieve without a security policy being enforced. 
Results for \texttt{Random, Bandwidth Weighted,} and \texttt{Inverse Bandwidth Weighted} deployments are found in \autoref{fig:security}.  We iterate over the total number of relays in the network, $p$, and quantify the number of circuits complying to each security policy.
These results analyze circuits generated for all security policies. 
Results for the \texttt{Position-Weighted} deployment are found in \autoref{fig:security2}. We iterate over weights $w_e$, $w_m$, and $w_x$, depending on the TEE distribution, and quantify the number of circuits complying to the distribution. These results only analyze circuits generated with the security policy synonymous with the deployment distribution. For example, the Entry-Exit distribution's security results only present circuits that had both a TEE-based entry relay and TEE-based exit relay.

\shortsection{Random}
Security results for \texttt{Random} deployment are in \autoref{security-rand}. 
These results show what security can be achieved with no strategic placement of TEEs in the network.

Statistically speaking, as $w$ is uniform across all relays, the probability of a circuit containing a TEE-based relay in a given position is equivalent to $p$. 
This is seen in the fact the percentage of TEE-based entry relay circuits is almost identical to the percentage of TEE-based exit relay circuits, which are both almost identical to $p$.

As expected, circuits with multiple TEE-based relays are less prevalent. Statistically, the probability both a TEE-based entry and TEE-based exit relay are in a circuit is $p^2$. 
Then, the probability of each position in the circuit containing a TEE-based relay is $p^3$.
Under this deployment, for circuits to contain a TEE-based relay in every position, at least an 80\% TEE presence is required.

\shortsection{Bandwidth Weighted}
Results for \texttt{Bandwidth Weighted} deployment are shown in \autoref{security-bw}.
This deployment provides the highest number of circuits with TEE-based relays under all policies in comparison to all deployments. There is an increased likelihood of a TEE-based relay being present because of Tor's relay selection algorithm, as the relays with more bandwidth are selected to be TEE-based. 

At only 23\% TEE presence, 50\% of circuits have a TEE-based entry relay, protecting users from two classes of attacks. At 28\% TEE presence, 50\% of circuits have a TEE-based exit relay, also protecting users from two classes of attacks. 
More significantly, though, at 44\% TEE presence, more than 50\% of circuits have a TEE-based relay in every position, providing protection from all classes of attacks.

\shortsection{Inverse Bandwidth Weighted}
The \texttt{Inverse Bandwidth Weighted} deployment is shown in \autoref{security-inversebw}. 
This deployment has the lowest number of circuits with TEE-based relays under all policies in comparison to all the deployments. 
This is expected, once again, because of Tor's relay selection algorithm. The likelihood of receiving a circuit with a TEE-based relay is significantly reduced in this deployment because of the minimal bandwidth they offer the network. 

With this deployment, in order for 50\% of circuits to have at least one TEE-based relay, a TEE penetration of greater than 80\% is required.
For comparison, this is the same deployment percentage required for more than 50\% of circuits to have all TEE-based relays (protection from every class of attack) in the \texttt{Random} deployment.

\input{figures/performance}

\shortsection{Circuit-Position Weighted}
The security results of \texttt{Circuit-\allowbreak{}Position Weighted} deployment provide insight into which types of relays adopting TEEs can most improve security.
Results for Entry distribution and Exit distribution are overlaid in \autoref{sec-overlay}. With this, the Entry distribution is iterating over $w_e$, while the Exit distribution is iterating over $w_x$.
The results of these distributions are defined in the same manner as \texttt{Random} deployment. For Entry, the probability of a circuit containing a TEE-based entry relay is equivalent to $w_e$. For Exit, the probability of a circuit containing a TEE-based exit relay is equivalent to $w_x$. With this, both deployment distributions grow linearly as $w_e$ and $w_x$ increase.

Entry-Exit (\autoref{sec-ex}) and Entry-Middle-Exit (\autoref{sec-emx-.25}, \autoref{sec-emx-.5}, and \autoref{sec-emx-.75}) continue with steady growth. Interestingly, at low values of $w_e$ but higher values of $w_x$, we see more circuits with TEE-based relays than one might expect. For example, with $w_e=1\%$ and $w_x=70\%$ in the Entry-Exit distribution, 10.9\% of circuits still had both a TEE-based entry relay and TEE-based exit relay. With the Entry-Middle-Exit distribution, adding in $w_m=50\%$, 29.5\% of circuits are entirely TEE-based. These results are because of the multiple capabilities relays may have. As all relays are middle-capable, and many exits are entry-capable, increasing $w_m$ and $w_x$ inherently increases $w_e$ in some manner.

These results also show the increase of TEE-based entry relays results in more circuits in comparison to the increase of TEE exit relays. For example, with $w_e=60\%$ and $w_x=30\%$ in the Entry-Exit distribution, 33.8\% of circuits have a TEE-based entry relay and TEE-based exit relay. With these values reversed, only 26.6\% of circuits have a TEE-based entry relay and TEE-based exit relay.

\shortsection{Takeaways}
Overall, \ParTEETor in non-policy mode provides users an immediate improvement to their security in terms of protection from known classes of attacks. 
While any TEE-based relay will benefit the network, higher bandwidth relays have the most influence because of how often they are used. Additionally, relays that are capable of being used in all positions of a circuit have a substantial impact, as they are not forced to adhere to one, or even two positions.
Therefore, relay operators with high bandwidth, or that allow for multiple uses are capable of significantly improving the security of Tor by adopting TEEs.

\input{figures/performance_2.tex}

\subsection{Performance of \ParTEETor in Policy Mode}

Recall that \ParTEETor in policy mode enforces security policies by extending the relay selection algorithm. 
This mode guarantees a user's security and is ideal for users concerned about protection from each class of attack. However, a potential consequence of this mode is increased congestion. Under limited deployments of TEEs, congestion may be more prevalent for users, as circuits will be forced to use the limited available TEE-based relays. 
To quantify this performance cost, we generate circuits with required TEE security policies. We then calculate median expected bandwidth of all circuits.

Results for \texttt{Random}, \texttt{Bandwidth Weighted}, and \texttt{Inverse Bandwidth Weighted} deployments are found in \autoref{fig:performance}.  We iterate over the total number of relays in the network, $p$.
These results analyze circuits generated for all security policies. 
Results for the \texttt{Position-Weighted} deployment are found in \autoref{fig:performance2}. We iterate over weights $w_e$, $w_m$, and $w_x$, depending on the TEE distribution. Once again, these results only analyze circuits generated with the security policy synonymous with the deployment distribution. In each figure, `baseline' represents the expected performance of users in Tor today, with no TEE requirements on circuits.
To normalize this value, we take the average across all the deployments, getting a final baseline expected bandwidth of 8347.3 KB/s.


\shortsection{Random}
Performance results for this deployment can be found in \autoref{perf-rand}.  
We see a steep increase immediately before tapering off with TEE-based entry security policy.  

Notably, at only 20\% TEEs in the network, circuits with a TEE-based entry relay requirement have a bandwidth of 7008.3 KB/s, a decrease of only 16\% from the baseline.
Considering more than half of the network is entry-capable, and entry relays have a minimum bandwidth they must meet, increasing $p$ inherently increases the number of entry relays at a faster rate than exit relays. Congestion subsides quickly for TEE-based entry circuits, allowing performance of users to be only minimally reduced.

In comparison, the TEE-based exit security policy has a more steady increase in performance, though bandwidths are substantially lower. This is because there are fewer exit relays in the network in comparison to entry, resulting in more congestion. At 20\% TEEs, circuits with a TEE-based exit relay have a 50.8\% decrease in performance from the baseline. 
We see with the final two security policies performance continues to degrade as more TEE-based relays are required in the circuits, as to be expected. All circuits are using the same few TEE-based relays, resulting in more congestion.

\shortsection{Bandwidth Weighted}
\sloppy The results for \texttt{Bandwidth Weighted} deployment 
 are shown in \autoref{perf-bw}.
This deployment
shows the highest performance in comparison to other deployments. 
All security policies have a dramatic increase in performance from 1\% TEE deployment until slowing around 25\% TEE deployment. 
By requiring TEE-based relays in a circuit, this deployment provides the highest performance because users are biased towards the highest bandwidth relays, as these relays are TEE-based. 

With only 10\% TEEs in the network, TEE-based entry circuits nearly meet the baseline bandwidth with only a 3\% decrease. 
At 44\% TEEs, all security policies have a bandwidth meeting and/or exceeding the baseline. With these results, users do not need to sacrifice their performance for increased security by requiring TEE-based relays in circuits.

The most compelling result from this deployment, though, is after 10\% TEEs in the network, TEE-based entry circuits surpass the baseline performance briefly, before slowing to baseline again around 70\% TEEs. Due to all the TEEs being concentrated on the highest-bandwidth relays, the relay selection algorithm is even more biased towards high performance, which we see at these stages (we evaluate the privacy implications of this in the next section).

\shortsection{Inverse Bandwidth Weighted}
We see a significant reduction in performance with \texttt{Inverse Bandwidth Weighted} deployment, seen in \autoref{perf-inversebw}. 
Circuits with the TEE-based entry security policy perform the best, as with all other deployments. This is, once again, due to the substantial number of entry relays in the network.
For TEE-based entry circuits to have at least half the bandwidth that is achieved by the baseline, 43\% TEEs is required. For comparison, at this same TEE percentage, full TEE-based circuits in the \texttt{Random} deployment exceed this bandwidth. More than a 70\% TEE presence is required for full TEE circuits to have this bandwidth.

Ultimately, this deployment scenario yields the most reduction in bandwidth compared to other deployments. This is to be expected, based on Tor's relay selection algorithm. As relays in this deployment have the least bandwidth in the network, requiring TEE-based relays, in turn, results in lower bandwidth.

\shortsection{Circuit-Position Weighted}
The \texttt{Circuit-Position Weighted} deployment provides insight into which types of relays are most critical to performance. 

The results for Entry distribution and Exit distribution are overlaid in \autoref{perf-overlay}. With this, the Entry distribution iterates over $w_e$, while the Exit distribution iterates over $w_x$.
Once again, these distributions reflect the results of the \texttt{Random} deployment. With 20\% TEE-based entry relays ($w_e$), circuits have a median bandwidth of 6941.7 KB/s, a 16.8\% decrease from baseline. 
This reflects the fact that more than half of the relays in the network are entry-capable, meaning congestion begins to taper off around 20\% TEEs.
For the same decrease in performance, 54\% of exit relays need to be TEE-based ($w_x$). 
A higher percentage of exit relays is required in comparison to entry relays because there are fewer exit relays in the network. Congestion is thus present at higher values of TEEs, as more circuits are vying for fewer existing TEE-based relays.

For the Entry-Exit distribution, shown in \autoref{perf-ex}, one might expect to see for low $w_e$ values similar results as in the previous two graphs. However, this is not the case, as we see when increasing the exit relays the bandwidth improves significantly, even with $w_e$ at only 1\%. This is because of the overlap in relays that are both entry and exit-capable. We see this with high $w_x$ values, the network naturally load balances itself by using the excess TEE-based exit relays as entry relays.
This same phenomenon is seen in the Entry-Middle-Exit distributions (\autoref{perf-emx-.25}, \autoref{perf-emx-.5}, \autoref{perf-emx-.75}). Despite the low values of $w_e$, as middle and exit relays gradually transition to adopting TEEs, the bandwidth of circuits increase because of the multi-capable relays in the network.

\input{figures/privacy}

\shortsection{Takeaways}
While one might expect to sacrifice performance to guarantee security with \ParTEETor in policy mode, our performance evaluation shows this is not always the case. When requiring TEE-based relays in circuits, congestion persists only until TEE percentages reach a critical mass point. High bandwidth relays contribute most to reaching this point at sparse TEE deployments, as they can withstand the demand of circuits requiring TEE-based relays. Relays that can function in multiple positions of a circuit also impact performance positively. As they can be used in various positions, they can help meet the demand of circuits requiring TEE-based relays when a specific type of relay is limited. 
With relay operators that provide high bandwidth or that allow for multiple uses adopting TEEs, the network can meet demands without significantly reducing user performance.

\subsection{Privacy of \ParTEETor in Policy Mode}

By specifying a security policy, the size of network is decreased, as circuits are limited to only containing TEE-based relays. Consequently, the space of available (policy-compliant) circuits is also reduced. When the space of possible circuits is reduced, the anonymity of users is impacted, as the predictability of circuits will be higher. Therefore, it is important to quantify the reduction in available circuits (and associated privacy) under specific TEE deployments and security policies. 
Through combinatorics, we derive the number of unique circuits that can be generated in \ParTEETor in policy mode for each security policy, in partial TEE deployments. We juxtapose these results against historical Tor data to justify why even the reduced space of available (TEE-based) circuits still provides reasonable privacy guarantees.

Recall that circuit position capabilities of relays are not exclusive. All relays are middle-capable, regardless
of their entry/exit capabilities.
Some relays are both entry and exit-capable.
We determine the breakdown of the number of relays with specific circuit position capabilities:
Middle-capable relays: $m_c = 6356$, Entry-capable relays: $e_c = 3179$, Exit-capable relays: $x_c = 1668$.
We denote \textit{p} as the total percentage of TEE-based relays in the network, representing our deployment scenario. We consider deployments of 1\%, 5\%, 10\%, 25\%, 50\%, and 75\%. For each, we consider a uniform deployment amongst each relay capability. 
The formulas derived are found in \autoref{tab:privacy}, along with our results. In baseline Tor, more than 33 billion unique circuits are possible.

Our results show that requiring only one relay in a circuit to be TEE-based, whether it be entry or exit, provides the least reduction in availability, as expected--these TEE security policies place the least restrictions on possible circuits, while also mitigating two classes of attacks. With a 1\% deployment, requiring a TEE-based entry relay results in only 1\% of the possible circuits in comparison to no TEE requirement; however, this still results in 336 million possible circuits. 
For comparison, this space of circuits is equivalent to that of Tor in May of 2010~\cite{tormetrics}. 
With a 25\% deployment of TEEs, the space of circuits with both a TEE-based entry relay and TEE-based exit relay is two billion, which is equivalent to Tor in March of 2012.
At a 50\% deployment of TEEs, the space of full TEE circuits is equivalent to Tor in July of 2013, with four billion circuits possible.



\shortsection{Takeaways}
Ultimately, our privacy results inform users on the trade-offs between security and privacy. By specifying a security policy, users
are guaranteed protection from known classes of attacks, and receive reasonable privacy guarantees. While the space of circuits is significantly reduced, it can still meet or exceed the expected privacy of historical versions of Tor used by hundreds of thousands of people. Increased TEE penetration will gradually resolve this issue.


%% file: figures/security.tex
\begin{figure*}[!t]
    \centering
    \captionsetup[subfigure]{font=scriptsize,labelfont=scriptsize}
        \begin{subfigure}[t]{.32\textwidth}
            \centering
            \includegraphics[width=\textwidth]{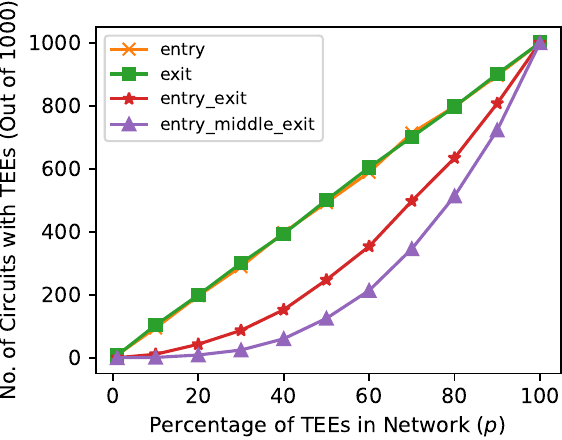}
            \caption{Random Relay}
            \label{security-rand}
        \end{subfigure}
        \hfill
        \begin{subfigure}[t]{.32\textwidth}
            \centering
            \includegraphics[width=\textwidth]{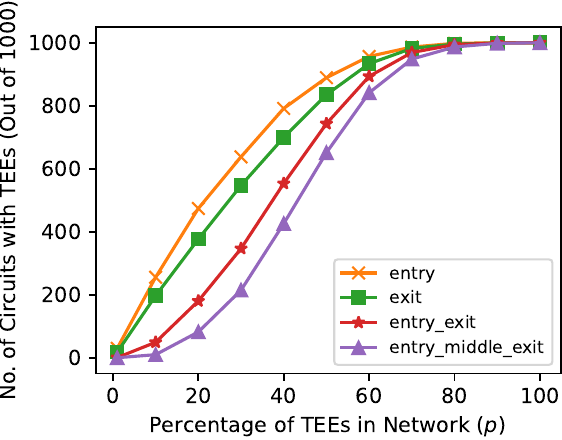}
            \caption{Bandwidth Weighted}
            \label{security-bw}
        \end{subfigure}
        \hfill
        \begin{subfigure}[t]{.32\textwidth}
            \centering
            \includegraphics[width=\textwidth]{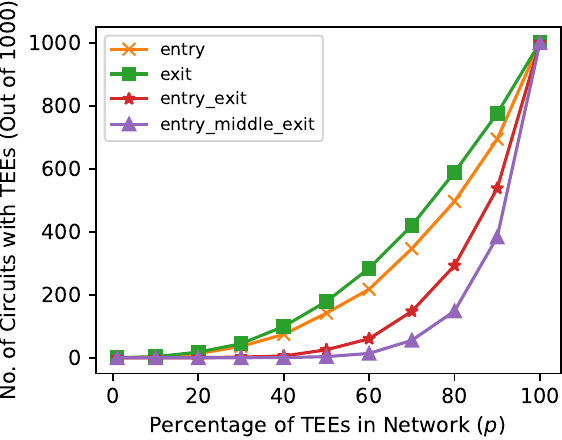}
            \caption{Inverse Bandwidth Weighted}
            \label{security-inversebw}
        \end{subfigure}

    \caption{TEE presence in circuits when no TEE security policy is specified when incrementing the percentage of TEEs present in the network, $p$. Results are the mean over 10 trials of 1000 circuits each.}
    \label{fig:security}
\end{figure*}

%% file: figures/security_2.tex
\begin{figure*}[!t]
    \captionsetup[subfigure]{font=scriptsize,labelfont=scriptsize}
    \centering
        \begin{subfigure}[t]{.32\textwidth}
            \centering
            \includegraphics[width=\textwidth]{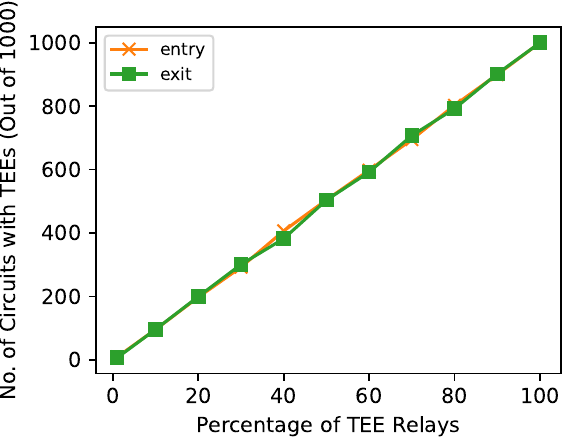}
            \caption{Entry Policy: $w_e=variable$; Exit Policy: $w_x=variable$}
            \label{sec-overlay}
        \end{subfigure}
        \hfill
        \hfill
        \begin{subfigure}[t]{.32\textwidth}
            \centering
            \includegraphics[width=\textwidth]{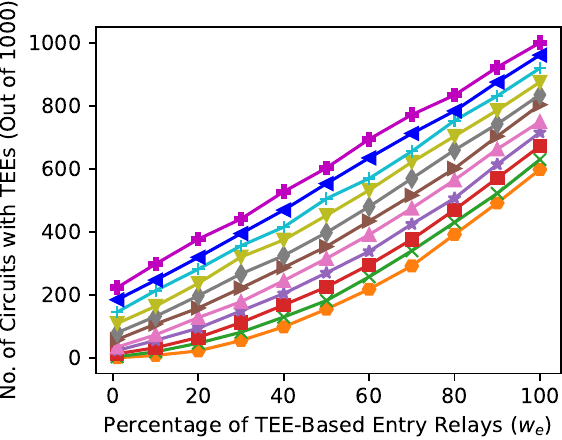}
            \caption{Entry-Exit Policy: $w_e=variable$, $w_x=variable$}
            \label{sec-ex}
        \end{subfigure}
        \begin{subfigure}[t]{.32\textwidth}
            \centering
            \includegraphics[width=\textwidth]{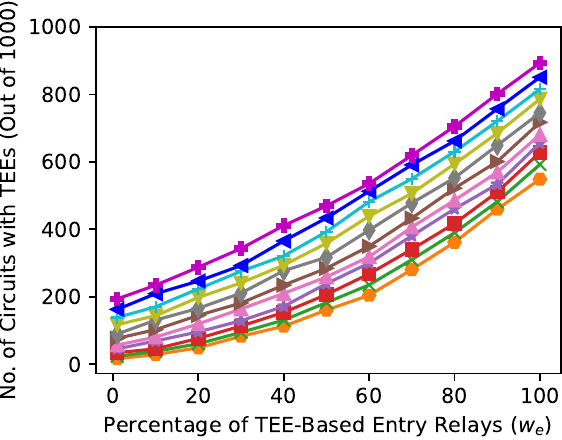}
            \caption{Entry-Middle-Exit Policy: $w_e=variable$, $w_m=25\%$, $w_x=variable$}
            \label{sec-emx-.25}
        \end{subfigure}
        \hfill
        \begin{subfigure}[t]{.32\textwidth}
            \centering
            \includegraphics[width=\textwidth]{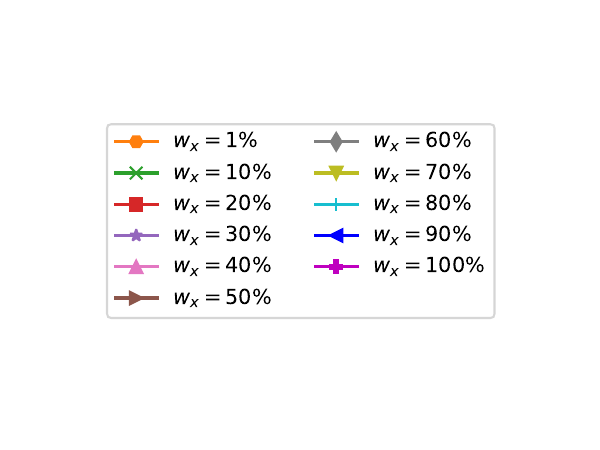}
            \caption{Legend for Entry-Exit and Entry-Middle-Exit distributions}
            \label{sec-legend}
        \end{subfigure}
        \hfill
        \begin{subfigure}[t]{.32\textwidth}
            \centering
            \includegraphics[width=\textwidth]{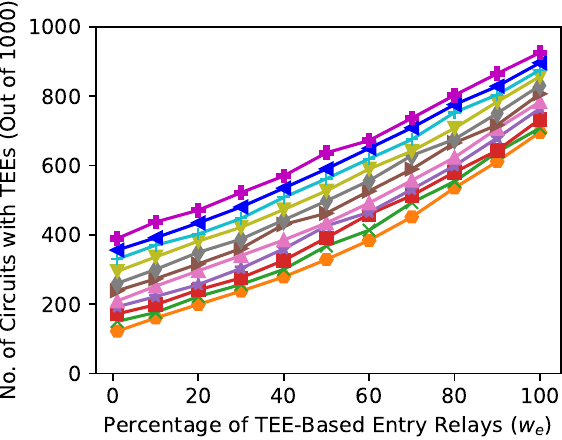}
            \caption{Entry-Middle-Exit Policy: $w_e=variable$, $w_m=50\%$, $w_x=variable$}
            \label{sec-emx-.5}
        \end{subfigure}
        \hfill
        \begin{subfigure}[t]{.32\textwidth}
            \centering
            \includegraphics[width=\textwidth]{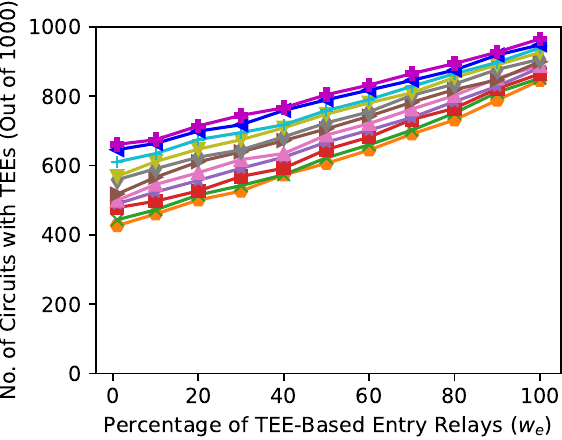}
            \caption{Entry-Middle-Exit Policy: $w_e=variable$, $w_m=75\%$, $w_x=variable$}
            \label{sec-emx-.75}
        \end{subfigure}
    \caption{TEE presence in circuits when no TEE security policy is specified for the \texttt{Circuit-Position Weighted} deployment. Each policy increments weights $w_e$, $w_m$, and $w_x$ as noted in the figures. Results are the mean over 10 trials  of 1000 circuits each.}
    \label{fig:security2}
\end{figure*}

%% file: figures/performance.tex
\begin{figure*}[!t]
    \centering
    \captionsetup[subfigure]{font=scriptsize,labelfont=scriptsize}
        \begin{subfigure}[t]{.31\textwidth}
            \centering
            \includegraphics[width=\textwidth]{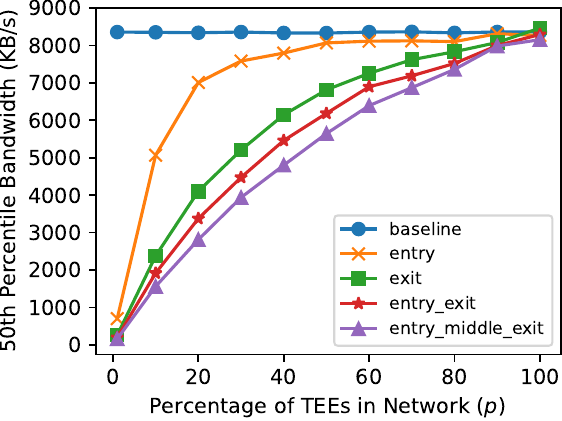}
            \caption{Random Relay}
            \label{perf-rand}
        \end{subfigure}
        \hfill
        \begin{subfigure}[t]{.31\textwidth}
            \centering
            \includegraphics[width=\textwidth]{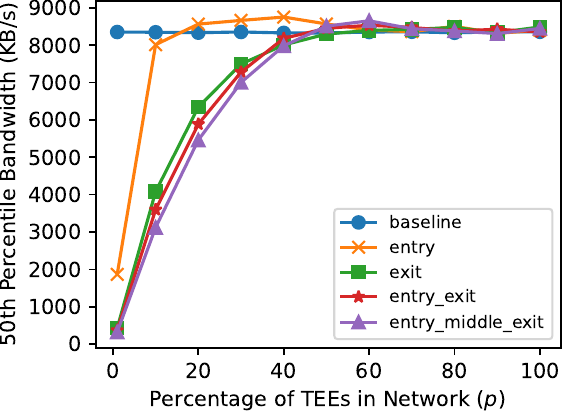}
            \caption{Bandwidth Weighted}
            \label{perf-bw}
        \end{subfigure}
        \hfill
        \begin{subfigure}[t]{.31\textwidth}
            \centering
            \includegraphics[width=\textwidth]{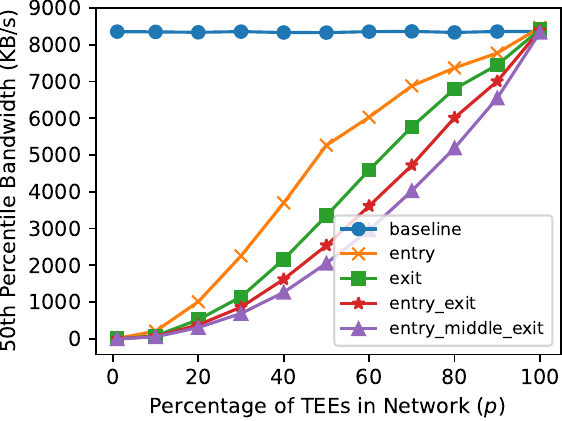}
            \caption{Inverse Bandwidth Weighted}
            \label{perf-inversebw}
        \end{subfigure}
    \caption{The median percentile bandwidth of circuits generated with each security policy  when incrementing the percentage of TEEs present in the network, $p$. Results are the mean over 10 trials of 1000 circuits each.}
    \label{fig:performance}
\end{figure*}

%% file: figures/performance_2.tex
\begin{figure*}[!t]
    \centering
    \captionsetup[subfigure]{font=scriptsize,labelfont=scriptsize}
        \begin{subfigure}[t]{.32\textwidth}
            \centering
            \includegraphics[width=\textwidth]{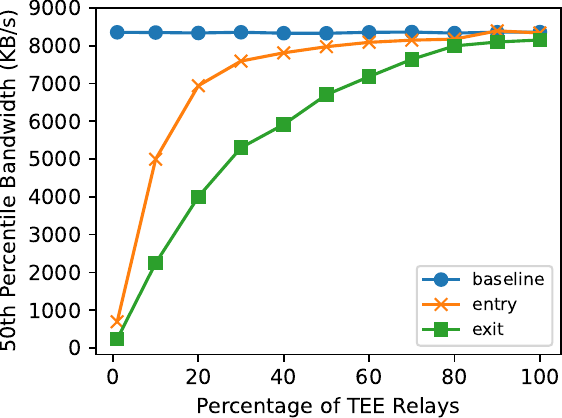}
            \caption{Entry Policy: $w_e=variable$; Exit Policy: $w_x=variable$}
            \label{perf-overlay}
        \end{subfigure}
        \hfill
        \hfill
        \begin{subfigure}[t]{.32\textwidth}
            \centering
            \includegraphics[width=\textwidth]{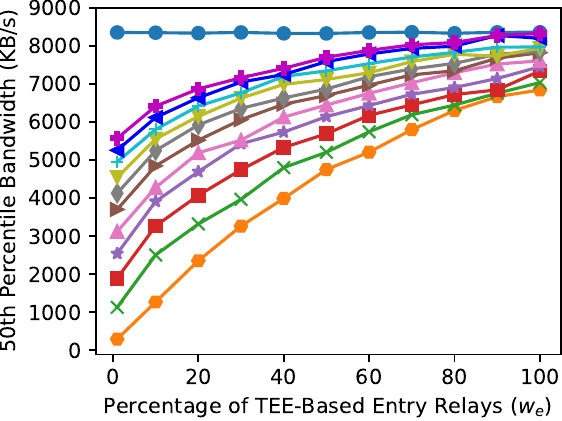}
            \caption{Entry-Exit Policy: $w_e=variable$, $w_x=variable$}
            \label{perf-ex}
        \end{subfigure}
        \begin{subfigure}[t]{.32\textwidth}
            \centering
            \includegraphics[width=\textwidth]{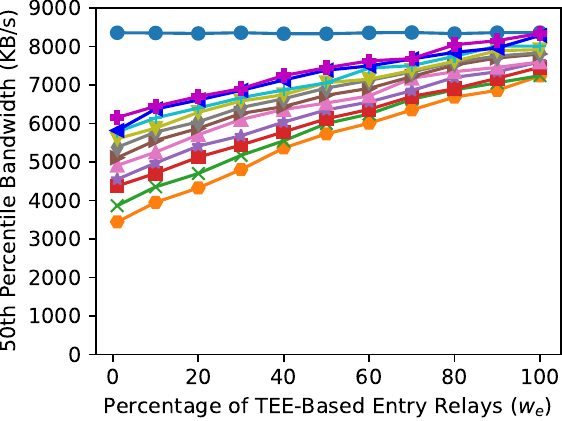}
            \caption{Entry-Middle-Exit Policy: $w_e=variable$, $w_m=25\%$, $w_x=variable$}
            \label{perf-emx-.25}
        \end{subfigure}
        \hfill
        \begin{subfigure}[t]{.32\textwidth}
            \centering
            \includegraphics[width=\textwidth]{./eval_graphs_shrunk/legend.pdf}
            \caption{Legend for Entry-Exit and Entry-Middle-Exit distributions}
            \label{perf-legend}
        \end{subfigure}
        \hfill
        \begin{subfigure}[t]{.32\textwidth}
            \centering
            \includegraphics[width=\textwidth]{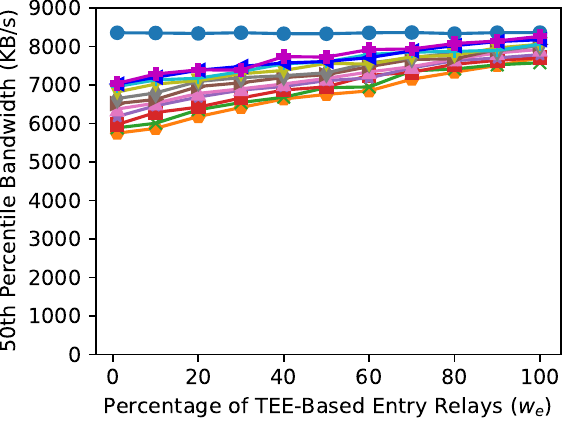}
            \caption{Entry-Middle-Exit Policy: $w_e=variable$, $w_m=50\%$, $w_x=variable$}
            \label{perf-emx-.5}
        \end{subfigure}
        \hfill
        \begin{subfigure}[t]{.32\textwidth}
            \centering
            \includegraphics[width=\textwidth]{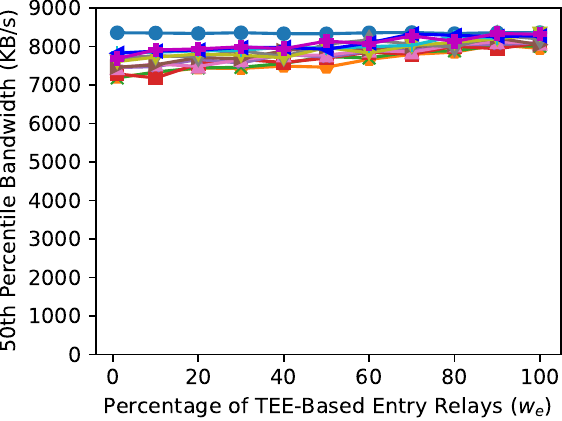}
            \caption{Entry-Middle-Exit Policy: $w_e=variable$, $w_m=75\%$, $w_x=variable$}
            \label{perf-emx-.75}
        \end{subfigure}
    \caption{The median percentile bandwidth of circuits generated for the \texttt{Circuit-Position Weighted} deployment, with each policy incrementing weights $w_e$, $w_m$, and $w_x$ as noted in the figures. Results are the mean over 10 trials of 1000 circuits each.}
    \label{fig:performance2}
\end{figure*}

%% file: figures/privacy.tex
\begin{table*}[!t]
    \centering
    \resizebox{\textwidth}{!}{
        \begin{tabular}{|l|l|l|l|l|l|l|l|}
            \hline
            \textbf{TEE Requirement} & \textbf{Number of Circuits}  & $\bm{p = 1\%}$  & $\bm{p = 5\%}$  & $\bm{p = 10\%}$  &  $\bm{p = 25\%}$ & $\bm{p = 50\%}$ & $\bm{p = 75\%}$ \\ \hline 
            None                   & $x_c(e_c - 1)(m_c - 2) $      & \multicolumn{6}{c|}{$3.36 \times 10^{10}$}         \\ \cline{3-8} 
            Entry                  & $x_c(pe_c - 1)(m_c - 2) $     & $3.36 \times 10^8$                 & $1.68 \times 10^9$        & $3.36 \times 10^9$  &  $8.42 \times 10^9$  & $1.68 \times 10^{10}$ & $2.52 \times 10^{10}$ \\
            Exit                   & $px_c(e_c - 1)(m_c - 2) $    & $3.36 \times 10^8$                 & $1.68 \times 10^9$             & $3.36 \times 10^9$  &   $8.42 \times 10^9$ & $1.68 \times 10^{10}$ & $2.52 \times 10^{10}$  \\
            Entry, Exit        & $px_c(pe_c - 1)(m_c - 2) $      & $3.36 \times 10^6$               & $8.42 \times 10^7$              & $3.36 \times 10^8$  &   $2.10 \times 10^9$ & $8.42 \times 10^9$ & $1.89 \times 10^{10}$ \\
            Entry, Middle, Exit & $px_c(pe_c - 1)(pm_c - 2) $    & $3.36 \times 10^4$          & $4.21 \times 10^6$                 & $3.36 \times 10^7$   &  $5.26 \times 10^8$ & $4.21 \times 10^9$ & $1.42 \times 10^{10}$  \\ \hline
        \end{tabular}
    }
    \caption{Number of unique circuits possible based on the ratio of TEE relays in the network and the security policy. $p$ represents the percentage of TEE relays in the network and $e_c$, $m_c$, and $x_c$ represent entry, middle, and exit capable relays, respectively.
    }
    \label{tab:privacy}
\end{table*}

%% file: conf-format/07.discussion.tex
\section{Discussion \& Related Work}\label{discussion}


\shortsection{Limitations}
While we demonstrate that TEEs are successful in defending against a broad class of attacks on Tor, we recognize their limitations: attacks against TEEs~\cite{gotzfried_cache_2017-1,lee_inferring_nodate,oleksenko_protecting_nodate,nilsson_survey_2020} and defenses~\cite{heiser2020towards,ahmad_obfuscuro_2019,sasy_zerotrace_2018} have been published.
Furthermore, TEEs can only reduce our adversary to a network level. The visibility that network administrators and ISPs have allows for more analysis on IP packets, which a local network adversary may not be able to analyze.

\shortsection{Relay Overloading} In this work, we do not consider relays reaching their bandwidth capacity. 
Consider the \texttt{Bandwidth-Weighted} deployment, for example. In this deployment, the highest performing relays are the most likely to be TEE-based. Due to Tor's relay selection algorithm, these relays are also most commonly used in circuits. This means that in this deployment, TEE-based relays will likely reach their bandwidth capacity because of their constant use. 
We leave to future work to model capacity limits of relays to better understand the impact on performance of users.

\shortsection{Implementing \ParTEETor in Tor}
We do not implement \ParTEETor in the actual Tor network. 
We acknowledge SGX-Tor's~\cite{sgx-tor-archival} technical challenges of implementing the Tor protocol in a TEE, which have been identified and addressed. For this reason, implementing \ParTEETor would largely reflect SGX-Tor, with only small modifications to the relay selection algorithm to support our security policies.

\shortsection{TEE-Augmented Tor} 
Other works have explored the use of TEEs for purposes largely unrelated to the attacks we evaluate here. For example, Panoply~\cite{panoply} considers in a case study implementing a directory authority within a TEE to to prevent adversaries from manipulating the consensus of the network. 
Jain \textit{et al.} propose OpenSGX~\cite{opensgx}, a system that emulates SGX at an instruction level. In their evaluation, they are motivated in preventing attacks that exploit the private keys of directory authorities and exit nodes, in which they integrate the cryptographic functions of the Tor protocol into their system. 
ConsenSGX~\cite{sasy_consensgx_2019}, considers the deployment of TEEs on directory cache relays for scalability problems. Clients are able to request only partial views of the network from TEE-enabled directory caches so they no longer need to store the entire consensus document. 
Note that none of these works consider the placement of the entire relay protocol within a TEE.

%% file: conf-format/09.conclusion.tex
\section{Conclusion}

In this paper we have demonstrated that partial deployments of TEE-based relays (and with small changes to the Tor operation) can substantially improve the resilience of the network to attacks.  Moreover, our analysis shows that such security gains grow with the increasing penetration of TEEs. 
Thus, we argue that integration of TEEs is not only an obvious win, but necessary to Tor's future.

%% file: main.bbl

\begin{thebibliography}{42}


\ifx \showCODEN    \undefined \def \showCODEN     #1{\unskip}     \fi
\ifx \showDOI      \undefined \def \showDOI       #1{#1}\fi
\ifx \showISBNx    \undefined \def \showISBNx     #1{\unskip}     \fi
\ifx \showISBNxiii \undefined \def \showISBNxiii  #1{\unskip}     \fi
\ifx \showISSN     \undefined \def \showISSN      #1{\unskip}     \fi
\ifx \showLCCN     \undefined \def \showLCCN      #1{\unskip}     \fi
\ifx \shownote     \undefined \def \shownote      #1{#1}          \fi
\ifx \showarticletitle \undefined \def \showarticletitle #1{#1}   \fi
\ifx \showURL      \undefined \def \showURL       {\relax}        \fi
\providecommand\bibfield[2]{#2}
\providecommand\bibinfo[2]{#2}
\providecommand\natexlab[1]{#1}
\providecommand\showeprint[2][]{arXiv:#2}

\bibitem[Ahmad et~al\mbox{.}(2019)]%
        {ahmad_obfuscuro_2019}
\bibfield{author}{\bibinfo{person}{Adil Ahmad}, \bibinfo{person}{Byunggill Joe}, \bibinfo{person}{Yuan Xiao}, \bibinfo{person}{Yinqian Zhang}, \bibinfo{person}{Insik Shin}, {and} \bibinfo{person}{Byoungyoung Lee}.} \bibinfo{year}{2019}\natexlab{}.
\newblock \showarticletitle{{OBFUSCURO}: {A} {Commodity} {Obfuscation} {Engine} on {Intel} {SGX}}. In \bibinfo{booktitle}{\emph{Proceedings 2019 {Network} and {Distributed} {System} {Security} {Symposium}}}. \bibinfo{publisher}{Internet Society}, \bibinfo{address}{San Diego, CA}, \bibinfo{numpages}{15}~pages.
\newblock
\showISBNx{978-1-891562-55-6}
\urldef\tempurl%
\url{https://doi.org/10.14722/ndss.2019.23513}
\showDOI{\tempurl}


\bibitem[AMD(2023)]%
        {sev_processors}
\bibfield{author}{\bibinfo{person}{AMD}.} \bibinfo{year}{2023}\natexlab{}.
\newblock \bibinfo{title}{Using SEV with AMD EPYC™ Processors}.
\newblock
\newblock
\urldef\tempurl%
\url{https://www.amd.com/content/dam/amd/en/documents/developer/58207-using-sev-with-amd-epyc-processors.pdf}
\showURL{%
\tempurl}


\bibitem[Apple(2024)]%
        {apple_enclave}
\bibfield{author}{\bibinfo{person}{Apple}.} \bibinfo{year}{2024}\natexlab{}.
\newblock \bibinfo{title}{Apple Platform Security: Secure {Enclave}}.
\newblock
\newblock
\urldef\tempurl%
\url{https://support.apple.com/guide/security/secure-enclave-sec59b0b31ff/web}
\showURL{%
\tempurl}


\bibitem[Bauer et~al\mbox{.}(2007)]%
        {bandwidth_attack}
\bibfield{author}{\bibinfo{person}{Kevin Bauer}, \bibinfo{person}{Damon McCoy}, \bibinfo{person}{Dirk Grunwald}, \bibinfo{person}{Tadayoshi Kohno}, {and} \bibinfo{person}{Douglas Sicker}.} \bibinfo{year}{2007}\natexlab{}.
\newblock \showarticletitle{Low-Resource Routing Attacks against Tor}. In \bibinfo{booktitle}{\emph{Proceedings of the 2007 ACM Workshop on Privacy in Electronic Society}} (Alexandria, Virginia, USA) \emph{(\bibinfo{series}{WPES '07})}. \bibinfo{publisher}{Association for Computing Machinery}, \bibinfo{address}{New York, NY, USA}, \bibinfo{pages}{11–20}.
\newblock
\showISBNx{9781595938831}
\urldef\tempurl%
\url{https://doi.org/10.1145/1314333.1314336}
\showDOI{\tempurl}


\bibitem[Biryukov et~al\mbox{.}(2013)]%
        {hidden_service_attack}
\bibfield{author}{\bibinfo{person}{Alex Biryukov}, \bibinfo{person}{Ivan Pustogarov}, {and} \bibinfo{person}{Ralf-Philipp Weinmann}.} \bibinfo{year}{2013}\natexlab{}.
\newblock \showarticletitle{Trawling for Tor Hidden Services: Detection, Measurement, Deanonymization}. In \bibinfo{booktitle}{\emph{2013 IEEE Symposium on Security and Privacy}}. \bibinfo{publisher}{IEEE Computer Society}, \bibinfo{address}{Oakland, California}, \bibinfo{pages}{80--94}.
\newblock
\urldef\tempurl%
\url{https://doi.org/10.1109/SP.2013.15}
\showDOI{\tempurl}


\bibitem[Blond et~al\mbox{.}(2011)]%
        {badapple_attack}
\bibfield{author}{\bibinfo{person}{Stevens~Le Blond}, \bibinfo{person}{Pere Manils}, \bibinfo{person}{Abdelberi Chaabane}, \bibinfo{person}{Mohamed~Ali Kaafar}, \bibinfo{person}{Claude Castelluccia}, \bibinfo{person}{Arnaud Legout}, {and} \bibinfo{person}{Walid Dabbous}.} \bibinfo{year}{2011}\natexlab{}.
\newblock \showarticletitle{One Bad Apple Spoils the Bunch: Exploiting {P2P} Applications to Trace and Profile Tor Users}. In \bibinfo{booktitle}{\emph{4th USENIX Workshop on Large-Scale Exploits and Emergent Threats (LEET 11)}}. \bibinfo{publisher}{USENIX Association}, \bibinfo{address}{Boston, MA}.
\newblock


\bibitem[che Tsai et~al\mbox{.}(2017)]%
        {graphene}
\bibfield{author}{\bibinfo{person}{Chia che Tsai}, \bibinfo{person}{Donald~E. Porter}, {and} \bibinfo{person}{Mona Vij}.} \bibinfo{year}{2017}\natexlab{}.
\newblock \showarticletitle{{Graphene-SGX}: A Practical Library {OS} for Unmodified Applications on {SGX}}. In \bibinfo{booktitle}{\emph{2017 USENIX Annual Technical Conference (USENIX ATC 17)}}. \bibinfo{publisher}{USENIX Association}, \bibinfo{address}{Santa Clara, CA}, \bibinfo{numpages}{14}~pages.
\newblock
\showISBNx{978-1-931971-38-6}


\bibitem[Danezis et~al\mbox{.}(2003)]%
        {danezis2003mixminion}
\bibfield{author}{\bibinfo{person}{George Danezis}, \bibinfo{person}{Roger Dingledine}, {and} \bibinfo{person}{Nick Mathewson}.} \bibinfo{year}{2003}\natexlab{}.
\newblock \showarticletitle{Mixminion: Design of a type III anonymous remailer protocol}. In \bibinfo{booktitle}{\emph{2003 Symposium on Security and Privacy}}. IEEE, \bibinfo{publisher}{IEEE}, \bibinfo{address}{Berkeley, CA}, \bibinfo{pages}{2--15}.
\newblock


\bibitem[Dingledine et~al\mbox{.}(2004)]%
        {tor2004original}
\bibfield{author}{\bibinfo{person}{Roger Dingledine}, \bibinfo{person}{Nick Mathewson}, {and} \bibinfo{person}{Paul Syverson}.} \bibinfo{year}{2004}\natexlab{}.
\newblock \showarticletitle{Tor: The {Second-Generation} Onion Router}. In \bibinfo{booktitle}{\emph{13th USENIX Security Symposium (USENIX Security 04)}}. \bibinfo{publisher}{USENIX Association}, \bibinfo{address}{San Diego, CA}.
\newblock


\bibitem[Evans et~al\mbox{.}(2009)]%
        {congestion_attack}
\bibfield{author}{\bibinfo{person}{Nathan~S. Evans}, \bibinfo{person}{Roger Dingledine}, {and} \bibinfo{person}{Christian Grothoff}.} \bibinfo{year}{2009}\natexlab{}.
\newblock \showarticletitle{A Practical Congestion Attack on Tor Using Long Paths}. In \bibinfo{booktitle}{\emph{Proceedings of the 18th Conference on USENIX Security Symposium}} (Montreal, Canada) \emph{(\bibinfo{series}{SSYM'09})}. \bibinfo{publisher}{USENIX Association}, \bibinfo{address}{USA}, \bibinfo{pages}{33–50}.
\newblock


\bibitem[Freedman and Morris(2002)]%
        {freedman2002tarzan}
\bibfield{author}{\bibinfo{person}{Michael~J Freedman} {and} \bibinfo{person}{Robert Morris}.} \bibinfo{year}{2002}\natexlab{}.
\newblock \showarticletitle{Tarzan: A peer-to-peer anonymizing network layer}. In \bibinfo{booktitle}{\emph{Proceedings of the 9th ACM Conference on Computer and Communications Security}}. \bibinfo{publisher}{ACM}, \bibinfo{address}{Washington, DC}, \bibinfo{pages}{193--206}.
\newblock


\bibitem[Goldschlag et~al\mbox{.}(1996)]%
        {onion_routing}
\bibfield{author}{\bibinfo{person}{David~M. Goldschlag}, \bibinfo{person}{Michael~G. Reed}, {and} \bibinfo{person}{Paul~F. Syverson}.} \bibinfo{year}{1996}\natexlab{}.
\newblock \showarticletitle{Hiding Routing Information}. In \bibinfo{booktitle}{\emph{Proceedings of the First International Workshop on Information Hiding}}. \bibinfo{publisher}{Springer-Verlag}, \bibinfo{address}{Berlin, Heidelberg}, \bibinfo{pages}{137–150}.
\newblock
\showISBNx{3540619968}


\bibitem[Götzfried et~al\mbox{.}(2017)]%
        {gotzfried_cache_2017-1}
\bibfield{author}{\bibinfo{person}{Johannes Götzfried}, \bibinfo{person}{Moritz Eckert}, \bibinfo{person}{Sebastian Schinzel}, {and} \bibinfo{person}{Tilo Müller}.} \bibinfo{year}{2017}\natexlab{}.
\newblock \showarticletitle{Cache {Attacks} on {Intel} {SGX}}. In \bibinfo{booktitle}{\emph{Proceedings of the 10th {European} {Workshop} on {Systems} {Security}}}. \bibinfo{publisher}{ACM}, \bibinfo{address}{Belgrade Serbia}, \bibinfo{pages}{1--6}.
\newblock
\showISBNx{978-1-4503-4935-2}
\urldef\tempurl%
\url{https://doi.org/10.1145/3065913.3065915}
\showDOI{\tempurl}


\bibitem[Heiser et~al\mbox{.}(2020)]%
        {heiser2020towards}
\bibfield{author}{\bibinfo{person}{Gernot Heiser}, \bibinfo{person}{Toby Murray}, {and} \bibinfo{person}{Gerwin Klein}.} \bibinfo{year}{2020}\natexlab{}.
\newblock \showarticletitle{Towards provable timing-channel prevention}.
\newblock \bibinfo{journal}{\emph{ACM SIGOPS Operating Systems Review}} \bibinfo{volume}{54}, \bibinfo{number}{1} (\bibinfo{year}{2020}), \bibinfo{pages}{1--7}.
\newblock


\bibitem[Intel(2015)]%
        {sgx}
\bibfield{author}{\bibinfo{person}{Intel}.} \bibinfo{year}{2015}\natexlab{}.
\newblock \bibinfo{title}{Intel® Software Guard Extensions}.
\newblock \bibinfo{howpublished}{\url{https://www.intel.com/content/www/us/en/developer/tools/software-guard-extensions/overview.html}}.
\newblock


\bibitem[Intel(2023)]%
        {sgx_processors}
\bibfield{author}{\bibinfo{person}{Intel}.} \bibinfo{year}{2023}\natexlab{}.
\newblock \bibinfo{title}{Intel® Processors Supporting Intel® SGX}.
\newblock
\newblock
\urldef\tempurl%
\url{https://www.intel.com/content/www/us/en/architecture-and-technology/software-guard-extensions-processors.html}
\showURL{%
\tempurl}


\bibitem[Jain et~al\mbox{.}(2016)]%
        {opensgx}
\bibfield{author}{\bibinfo{person}{Prerit Jain}, \bibinfo{person}{Soham Desai}, \bibinfo{person}{Seongmin Kim}, \bibinfo{person}{Ming-Wei Shih}, \bibinfo{person}{JaeHyuk Lee}, \bibinfo{person}{Changho Choi}, \bibinfo{person}{Youjung Shin}, \bibinfo{person}{Taesoo Kim}, \bibinfo{person}{Brent Byunghoon~Kang}, {and} \bibinfo{person}{Dongsu Han}.} \bibinfo{year}{2016}\natexlab{}.
\newblock \showarticletitle{{OpenSGX}: {An} {Open} {Platform} for {SGX} {Research}}. In \bibinfo{booktitle}{\emph{Proceedings 2016 {Network} and {Distributed} {System} {Security} {Symposium}}}. \bibinfo{publisher}{Internet Society}, \bibinfo{address}{San Diego, CA}.
\newblock
\showISBNx{978-1-891562-41-9}
\urldef\tempurl%
\url{https://doi.org/10.14722/ndss.2016.23011}
\showDOI{\tempurl}


\bibitem[Jansen et~al\mbox{.}(2012)]%
        {jansen2012methodically}
\bibfield{author}{\bibinfo{person}{Rob Jansen}, \bibinfo{person}{Kevin~S Bauer}, \bibinfo{person}{Nicholas Hopper}, {and} \bibinfo{person}{Roger Dingledine}.} \bibinfo{year}{2012}\natexlab{}.
\newblock \showarticletitle{Methodically Modeling the Tor Network}. In \bibinfo{booktitle}{\emph{5th Workshop on Cyber Security Experimentation and Test (CSET 12)}}. \bibinfo{publisher}{USENIX Association}, \bibinfo{address}{Bellevue, WA}.
\newblock


\bibitem[Jansen et~al\mbox{.}(2014)]%
        {jansen_sniper_2014}
\bibfield{author}{\bibinfo{person}{Rob Jansen}, \bibinfo{person}{Florian Tschorsch}, \bibinfo{person}{Aaron Johnson}, {and} \bibinfo{person}{Björn Scheuermann}.} \bibinfo{year}{2014}\natexlab{}.
\newblock \showarticletitle{The {Sniper} {Attack}: {Anonymously} {Deanonymizing} and {Disabling} the {Tor} {Network}}. In \bibinfo{booktitle}{\emph{Proceedings 2014 {Network} and {Distributed} {System} {Security} {Symposium}}}. \bibinfo{publisher}{Internet Society}, \bibinfo{address}{San Diego, CA}.
\newblock
\showISBNx{978-1-891562-35-8}
\urldef\tempurl%
\url{https://doi.org/10.14722/ndss.2014.23288}
\showDOI{\tempurl}


\bibitem[Johnson(2022)]%
        {stempython}
\bibfield{author}{\bibinfo{person}{Damian Johnson}.} \bibinfo{year}{2022}\natexlab{}.
\newblock \bibinfo{title}{Directory — {Stem} 1.8.0 documentation}.
\newblock \bibinfo{howpublished}{\url{https://stem.torproject.org/api/directory.html}}.
\newblock


\bibitem[(juga)(2019)]%
        {bandwidth_scanner}
\bibfield{author}{\bibinfo{person}{Tor~Blog (juga)}.} \bibinfo{year}{2019}\natexlab{}.
\newblock \bibinfo{title}{How Bandwidth Scanners Monitor The Tor Network}.
\newblock \bibinfo{howpublished}{\url{https://blog.torproject.org/how-bandwidth-scanners-monitor-tor-network/}}.
\newblock


\bibitem[Kaplan et~al\mbox{.}(2016)]%
        {kaplan_amd_2016}
\bibfield{author}{\bibinfo{person}{David Kaplan}, \bibinfo{person}{Jeremy Powell}, {and} \bibinfo{person}{Tom Woller}.} \bibinfo{year}{2016}\natexlab{}.
\newblock \showarticletitle{AMD memory encryption}.
\newblock \bibinfo{journal}{\emph{White paper}}  \bibinfo{volume}{13} (\bibinfo{year}{2016}), \bibinfo{numpages}{12}~pages.
\newblock


\bibitem[Kim et~al\mbox{.}(2018)]%
        {sgx-tor-archival}
\bibfield{author}{\bibinfo{person}{Seongmin Kim}, \bibinfo{person}{Juhyeng Han}, \bibinfo{person}{Jaehyeong Ha}, \bibinfo{person}{Taesoo Kim}, {and} \bibinfo{person}{Dongsu Han}.} \bibinfo{year}{2018}\natexlab{}.
\newblock \showarticletitle{SGX-Tor: A Secure and Practical Tor Anonymity Network With SGX Enclaves}.
\newblock \bibinfo{journal}{\emph{IEEE/ACM Transactions on Networking}} \bibinfo{volume}{26}, \bibinfo{number}{5} (\bibinfo{year}{2018}), \bibinfo{pages}{2174--2187}.
\newblock
\urldef\tempurl%
\url{https://doi.org/10.1109/TNET.2018.2868054}
\showDOI{\tempurl}


\bibitem[Kwon et~al\mbox{.}(2015)]%
        {fingerprinting_attack}
\bibfield{author}{\bibinfo{person}{Albert Kwon}, \bibinfo{person}{Mashael AlSabah}, \bibinfo{person}{David Lazar}, \bibinfo{person}{Marc Dacier}, {and} \bibinfo{person}{Srinivas Devadas}.} \bibinfo{year}{2015}\natexlab{}.
\newblock \showarticletitle{Circuit Fingerprinting Attacks: Passive Deanonymization of Tor Hidden Services}. In \bibinfo{booktitle}{\emph{24th USENIX Security Symposium (USENIX Security 15)}}. \bibinfo{publisher}{USENIX Association}, \bibinfo{address}{Washington, D.C.}, \bibinfo{pages}{287--302}.
\newblock
\showISBNx{978-1-939133-11-3}


\bibitem[Lee et~al\mbox{.}(2017)]%
        {lee_inferring_nodate}
\bibfield{author}{\bibinfo{person}{Sangho Lee}, \bibinfo{person}{Ming-Wei Shih}, \bibinfo{person}{Prasun Gera}, \bibinfo{person}{Taesoo Kim}, \bibinfo{person}{Hyesoon Kim}, {and} \bibinfo{person}{Marcus Peinado}.} \bibinfo{year}{2017}\natexlab{}.
\newblock \showarticletitle{Inferring Fine-grained Control Flow Inside {SGX} Enclaves with Branch Shadowing}. In \bibinfo{booktitle}{\emph{26th USENIX Security Symposium (USENIX Security 17)}}. \bibinfo{publisher}{USENIX Association}, \bibinfo{address}{Vancouver, BC}, \bibinfo{pages}{557--574}.
\newblock
\showISBNx{978-1-931971-40-9}
\urldef\tempurl%
\url{https://www.usenix.org/conference/usenixsecurity17/technical-sessions/presentation/lee-sangho}
\showURL{%
\tempurl}


\bibitem[Limited(2009)]%
        {trustzone2009}
\bibfield{author}{\bibinfo{person}{Arm Limited}.} \bibinfo{year}{2009}\natexlab{}.
\newblock \bibinfo{title}{ARM Security Technology Building a Secure System using TrustZone Technology}.
\newblock
\newblock
\urldef\tempurl%
\url{https://developer.arm.com/documentation/PRD29-GENC-009492/latest/}
\showURL{%
\tempurl}


\bibitem[Limited(2017)]%
        {trustzone2017}
\bibfield{author}{\bibinfo{person}{Arm Limited}.} \bibinfo{year}{2017}\natexlab{}.
\newblock \bibinfo{title}{TrustZone technology for ARMv8-M Architecture Version 2.1}.
\newblock
\newblock
\urldef\tempurl%
\url{https://developer.arm.com/documentation/100690/latest/}
\showURL{%
\tempurl}


\bibitem[McKeen et~al\mbox{.}(2013)]%
        {mckeen2013innovative}
\bibfield{author}{\bibinfo{person}{Frank McKeen}, \bibinfo{person}{Ilya Alexandrovich}, \bibinfo{person}{Alex Berenzon}, \bibinfo{person}{Carlos~V Rozas}, \bibinfo{person}{Hisham Shafi}, \bibinfo{person}{Vedvyas Shanbhogue}, {and} \bibinfo{person}{Uday~R Savagaonkar}.} \bibinfo{year}{2013}\natexlab{}.
\newblock \showarticletitle{Innovative instructions and software model for isolated execution.}
\newblock \bibinfo{journal}{\emph{Hasp@ isca}} \bibinfo{volume}{10}, \bibinfo{number}{1} (\bibinfo{year}{2013}), \bibinfo{numpages}{8}~pages.
\newblock


\bibitem[Ngabonziza et~al\mbox{.}(2016)]%
        {ngabonziza_trustzone_2016}
\bibfield{author}{\bibinfo{person}{Bernard Ngabonziza}, \bibinfo{person}{Daniel Martin}, \bibinfo{person}{Anna Bailey}, \bibinfo{person}{Haehyun Cho}, {and} \bibinfo{person}{Sarah Martin}.} \bibinfo{year}{2016}\natexlab{}.
\newblock \showarticletitle{{TrustZone} {Explained}: {Architectural} {Features} and {Use} {Cases}}. In \bibinfo{booktitle}{\emph{2016 {IEEE} 2nd {International} {Conference} on {Collaboration} and {Internet} {Computing} ({CIC})}}. \bibinfo{publisher}{IEEE}, \bibinfo{address}{Pittsburgh, PA, USA}, \bibinfo{pages}{445--451}.
\newblock
\showISBNx{978-1-5090-4607-2}
\urldef\tempurl%
\url{https://doi.org/10.1109/CIC.2016.065}
\showDOI{\tempurl}


\bibitem[Nilsson et~al\mbox{.}(2020)]%
        {nilsson_survey_2020}
\bibfield{author}{\bibinfo{person}{Alexander Nilsson}, \bibinfo{person}{Pegah~Nikbakht Bideh}, {and} \bibinfo{person}{Joakim Brorsson}.} \bibinfo{year}{2020}\natexlab{}.
\newblock \showarticletitle{A {Survey} of {Published} {Attacks} on {Intel} {SGX}}.
\newblock \bibinfo{journal}{\emph{CoRR}} (\bibinfo{date}{June} \bibinfo{year}{2020}), \bibinfo{numpages}{11}~pages.
\newblock
\urldef\tempurl%
\url{http://arxiv.org/abs/2006.13598}
\showURL{%
\tempurl}
\newblock
\shownote{arXiv:2006.13598 [cs]}.


\bibitem[Oleksenko et~al\mbox{.}(2018)]%
        {oleksenko_protecting_nodate}
\bibfield{author}{\bibinfo{person}{Oleksii Oleksenko}, \bibinfo{person}{Bohdan Trach}, \bibinfo{person}{Robert Krahn}, \bibinfo{person}{Mark Silberstein}, {and} \bibinfo{person}{Christof Fetzer}.} \bibinfo{year}{2018}\natexlab{}.
\newblock \showarticletitle{Varys: Protecting {SGX} Enclaves from Practical {Side-Channel} Attacks}. In \bibinfo{booktitle}{\emph{2018 USENIX Annual Technical Conference (USENIX ATC 18)}}. \bibinfo{publisher}{USENIX Association}, \bibinfo{address}{Boston, MA}, \bibinfo{pages}{227--240}.
\newblock
\showISBNx{ISBN 978-1-939133-01-4}
\urldef\tempurl%
\url{https://www.usenix.org/conference/atc18/presentation/oleksenko}
\showURL{%
\tempurl}


\bibitem[Overlier and Syverson(2006)]%
        {demultiplex_attack}
\bibfield{author}{\bibinfo{person}{L. Overlier} {and} \bibinfo{person}{P. Syverson}.} \bibinfo{year}{2006}\natexlab{}.
\newblock \showarticletitle{Locating hidden servers}. In \bibinfo{booktitle}{\emph{2006 IEEE Symposium on Security and Privacy (S P'06)}}. \bibinfo{publisher}{IEEE Computer Society}, \bibinfo{address}{Oakland, California}, \bibinfo{pages}{15 pp.--114}.
\newblock
\urldef\tempurl%
\url{https://doi.org/10.1109/SP.2006.24}
\showDOI{\tempurl}


\bibitem[Panchenko et~al\mbox{.}(2012)]%
        {panchenko_improving_2012}
\bibfield{author}{\bibinfo{person}{Andriy Panchenko}, \bibinfo{person}{Fabian Lanze}, {and} \bibinfo{person}{Thomas Engel}.} \bibinfo{year}{2012}\natexlab{}.
\newblock \showarticletitle{Improving performance and anonymity in the {Tor} network}. In \bibinfo{booktitle}{\emph{2012 {IEEE} 31st {International} {Performance} {Computing} and {Communications} {Conference} ({IPCCC})}}. \bibinfo{publisher}{IEEE}, \bibinfo{address}{Austin, TX, USA}, \bibinfo{pages}{1--10}.
\newblock
\showISBNx{978-1-4673-4883-6 978-1-4673-4881-2 978-1-4673-4882-9}
\urldef\tempurl%
\url{https://doi.org/10.1109/PCCC.2012.6407715}
\showDOI{\tempurl}


\bibitem[Pries et~al\mbox{.}(2008)]%
        {replay_tagging_attack}
\bibfield{author}{\bibinfo{person}{R. Pries}, \bibinfo{person}{W. Yu}, \bibinfo{person}{X. Fu}, {and} \bibinfo{person}{W. Zhao}.} \bibinfo{year}{2008}\natexlab{}.
\newblock \showarticletitle{A {New} {Replay} {Attack} {Against} {Anonymous} {Communication} {Networks}}. In \bibinfo{booktitle}{\emph{2008 {IEEE} {International} {Conference} on {Communications}}}. \bibinfo{publisher}{IEEE}, \bibinfo{address}{Beijing, China}, \bibinfo{pages}{1578--1582}.
\newblock
\showISBNx{978-1-4244-2075-9}
\urldef\tempurl%
\url{https://doi.org/10.1109/ICC.2008.305}
\showDOI{\tempurl}


\bibitem[Project(2022)]%
        {tormetrics}
\bibfield{author}{\bibinfo{person}{Tor Project}.} \bibinfo{year}{2022}\natexlab{}.
\newblock \bibinfo{title}{{Tor} {Metrics}}.
\newblock \bibinfo{howpublished}{\url{https://metrics.torproject.org/}}.
\newblock


\bibitem[Reiter and Rubin(1998)]%
        {reiter1998crowds}
\bibfield{author}{\bibinfo{person}{Michael~K Reiter} {and} \bibinfo{person}{Aviel~D Rubin}.} \bibinfo{year}{1998}\natexlab{}.
\newblock \showarticletitle{Crowds: Anonymity for web transactions}.
\newblock \bibinfo{journal}{\emph{ACM transactions on information and system security (TISSEC)}} \bibinfo{volume}{1}, \bibinfo{number}{1} (\bibinfo{year}{1998}), \bibinfo{pages}{66--92}.
\newblock


\bibitem[Sasy and Goldberg(2019)]%
        {sasy_consensgx_2019}
\bibfield{author}{\bibinfo{person}{Sajin Sasy} {and} \bibinfo{person}{Ian Goldberg}.} \bibinfo{year}{2019}\natexlab{}.
\newblock \showarticletitle{{ConsenSGX}: {Scaling} {Anonymous} {Communications} {Networks} with {Trusted} {Execution} {Environments}}. In \bibinfo{booktitle}{\emph{Proceedings on Privacy Enhancing Technologies}}. \bibinfo{publisher}{De Gruyter Open}, \bibinfo{address}{Stockholm, Sweden}, \bibinfo{pages}{331--349}.
\newblock
\showISSN{2299-0984}
\urldef\tempurl%
\url{https://doi.org/10.2478/popets-2019-0050}
\showDOI{\tempurl}


\bibitem[Sasy et~al\mbox{.}(2018)]%
        {sasy_zerotrace_2018}
\bibfield{author}{\bibinfo{person}{Sajin Sasy}, \bibinfo{person}{Sergey Gorbunov}, {and} \bibinfo{person}{Christopher~W. Fletcher}.} \bibinfo{year}{2018}\natexlab{}.
\newblock \showarticletitle{{ZeroTrace} : {Oblivious} {Memory} {Primitives} from {Intel} {SGX}}. In \bibinfo{booktitle}{\emph{Proceedings 2018 {Network} and {Distributed} {System} {Security} {Symposium}}}. \bibinfo{publisher}{Internet Society}, \bibinfo{address}{San Diego, CA}, \bibinfo{numpages}{15}~pages.
\newblock
\showISBNx{978-1-891562-49-5}
\urldef\tempurl%
\url{https://doi.org/10.14722/ndss.2018.23239}
\showDOI{\tempurl}


\bibitem[Shinde et~al\mbox{.}(2017)]%
        {panoply}
\bibfield{author}{\bibinfo{person}{Shweta Shinde}, \bibinfo{person}{Dat Le}, \bibinfo{person}{Shruti Tople}, {and} \bibinfo{person}{Prateek Saxena}.} \bibinfo{year}{2017}\natexlab{}.
\newblock \showarticletitle{Panoply: Low-TCB Linux Applications with SGX Enclaves}. In \bibinfo{booktitle}{\emph{24th Annual Network and Distributed System Security Symposium}}. \bibinfo{publisher}{Internet Society}, \bibinfo{address}{San Diego, California}.
\newblock
\urldef\tempurl%
\url{https://doi.org/10.14722/ndss.2017.23500}
\showDOI{\tempurl}


\bibitem[Snader and Borisov(2008)]%
        {tuneup}
\bibfield{author}{\bibinfo{person}{Robin Snader} {and} \bibinfo{person}{Nikita Borisov}.} \bibinfo{year}{2008}\natexlab{}.
\newblock \showarticletitle{A Tune-up for Tor: Improving Security and Performance in the Tor Network}. In \bibinfo{booktitle}{\emph{Proceedings 2008 {Network} and {Distributed} {System} {Security} {Symposium}}}, Vol.~\bibinfo{volume}{8}. \bibinfo{publisher}{Internet Society}, \bibinfo{address}{San Diego, California}, \bibinfo{pages}{127}.
\newblock


\bibitem[torproject(2022)]%
        {torspec_git}
\bibfield{author}{\bibinfo{person}{torproject}.} \bibinfo{year}{2022}\natexlab{}.
\newblock \bibinfo{title}{Tor Specifications}.
\newblock \bibinfo{howpublished}{\url{https://github.com/torproject/torspec}}.
\newblock


\bibitem[Zantout et~al\mbox{.}(2011)]%
        {zantout2011i2p}
\bibfield{author}{\bibinfo{person}{Bassam Zantout}, \bibinfo{person}{Ramzi Haraty}, {et~al\mbox{.}}} \bibinfo{year}{2011}\natexlab{}.
\newblock \showarticletitle{I2P data communication system}. In \bibinfo{booktitle}{\emph{The Tenth International Conference on Networks (ICN 2011)}}. \bibinfo{publisher}{ICN, Springer Singapore}, \bibinfo{address}{The Netherlands}, \bibinfo{pages}{401--409}.
\newblock


\end{thebibliography}
